\documentclass[manuscript,screen]{acmart}
\usepackage{booktabs}
\usepackage{multirow}
\usepackage{listings} 
\usepackage{tabularx}
\usepackage{graphicx} 
\usepackage{hyperref}
\usepackage{enumitem}
\usepackage{subcaption}
\usepackage{colortbl}

\setlist[itemize]{leftmargin=*} 
\AtBeginDocument{%
	}

\acmJournal{TOIS}
\begin{document}
	
	\title{KGBridge: Knowledge-Guided Prompt Learning for Non-overlapping Cross-Domain Recommendation}
	
	\author{Yuhan Wang}
	\orcid{0000-0003-1526-276X}
	\email{wyh0520@whut.edu.cn}
	\affiliation{%
		\institution{Wuhan University of Technology}
		\city{Wuhan}
		\country{China}
		\postcode{430070}}
	\affiliation{%
		\institution{Engineering Research Center of Intelligent Service Technology for Digital Publishing, Ministry of Education}
		\city{Wuhan}
		\country{China}
		\postcode{430070}}
	
	\author{Qing Xie}
	\orcid{0000-0003-4530-588X}
	\email{felixxq@whut.edu.cn}
	\authornote{Corresponding authors}
	\affiliation{%
		\institution{Wuhan University of Technology}
		\city{Wuhan}
		\country{China}
		\postcode{430070}}
	\affiliation{%
		\institution{Engineering Research Center of Intelligent Service Technology for Digital Publishing, Ministry of Education}
		\city{Wuhan}
		\country{China}
		\postcode{430070}}
	
	\author{Zhifeng Bao}
	\orcid{0000-0003-2477-381X}
	\email{zhifeng.bao@uq.edu.au}
	\affiliation{%
		\institution{The University of Queensland}
		\city{Brisbane}
		\country{Australia}
		\postcode{4000}}
	
	\author{Mengzi Tang}
	\orcid{0000-0001-5611-673X}
	\email{tangmz@whut.edu.cn}
	\affiliation{%
		\institution{Wuhan University of Technology}
		\city{Wuhan}
		\country{China}
		\postcode{430070}}
	\affiliation{%
		\institution{Engineering Research Center of Intelligent Service Technology for Digital Publishing, Ministry of Education}
		\city{Wuhan}
		\country{China}
		\postcode{430070}}
	
	\author{Lin Li}
	\orcid{0000-0001-7553-6916}
	\email{cathylilin@whut.edu.cn}
	\affiliation{%
		\institution{Wuhan University of Technology}
		\city{Wuhan}
		\country{China}
		\postcode{430070}}
	\affiliation{%
		\institution{Engineering Research Center of Intelligent Service Technology for Digital Publishing, Ministry of Education}
		\city{Wuhan}
		\country{China}
		\postcode{430070}}
	
	\author{Yongjian Liu}
	\orcid{0009-0005-8388-6736}
	\email{liuyj@whut.edu.cn}
	
	\affiliation{%
		\institution{Wuhan University of Technology}
		\city{Wuhan}
		\country{China}
		\postcode{430070}}
	\affiliation{%
		\institution{Engineering Research Center of Intelligent Service Technology for Digital Publishing, Ministry of Education}
		\city{Wuhan}
		\country{China}
		\postcode{430070}}
	
	\renewcommand{\shortauthors}{Wang et al.}
	
	\begin{abstract}
		Knowledge Graphs (KGs) provide structured relational information, offering a unified semantic foundation for cross-domain recommendation (CDR). Integrating symbolic knowledge with interactions enriches representations, supports reasoning, and enhances interpretability. However, existing methods face three major challenges: (\textbf{C1}) sensitivity to KG sparsity and popularity bias, (\textbf{C2}) reliance on overlapping users for domain alignment, and (\textbf{C3}) insufficient disentanglement between transferable and domain-specific knowledge, which limits stable knowledge transfer.
		
		To address these challenges, this paper proposes KGBridge, a knowledge-guided prompt learning framework for non-overlapping cross-domain sequential recommendation. KGBridge comprises: (i) a KG-enhanced prompt encoder that models relation-level semantics as soft prompts, providing structured priors for sequential preference modeling (for \textbf{C1}), and (ii) a two-stage training paradigm combining cross-domain pretraining with privacy-preserving fine-tuning to enable knowledge transfer without shared users (for \textbf{C2}). Furthermore, via correspondence-driven disentanglement, KGBridge explicitly separates domain-shared and domain-specific semantics, preserving their complementarity and stabilizing target-domain adaptation (for \textbf{C3}). Extensive experiments on benchmark datasets demonstrate that KGBridge consistently outperforms state-of-the-art baselines and exhibits robustness under varying KG sparsity, confirming its effectiveness in alleviating structural imbalance and semantic entanglement. To facilitate reproducibility, the source code and datasets are publicly available at \url{https://github.com/WangYuhan-0520/KGBridge}.
	\end{abstract}

	
	
	\maketitle
	\section{Introduction}
	Knowledge Graphs (KGs) have been widely adopted in recommender systems as structured knowledge sources that explicitly encode entities, relations, and their semantic dependencies \cite{GuoZQZXXH22, JiPCMY22, Dong23}. By organizing information into interconnected graph structures, KGs provide rich semantic priors that complement sparse user-item interaction data, thereby enhancing preference modeling, relational reasoning, and interpretability \cite{GuoZQZXXH22}. Numerous studies integrate KGs into recommendation models, typically by learning item-enriched representations and propagating relational information to capture high-order dependencies beyond direct interactions \cite{TangGW0WFZ24, BalloccuBFM22, LiuYZC24, WangS0WYK23, WangZWZLXG18, Wang00LC19, HuangZDWC18, CuiW0LL0025}. Notably, most existing KG-based methods primarily focus on single-domain settings, which limits the exploitation of inherently cross-domain semantics encoded in KGs \cite{GuoZQZXXH22, ZhuW00L021}. 
	
	Beyond single-domain recommendation, the relational structure of KGs naturally spans multiple domains via shared entities, attributes, and inter-domain relations, providing a unified semantic foundation for knowledge transfer across heterogeneous environments. This intrinsic property makes KGs particularly well-suited for cross-domain recommendation (CDR), which aims to alleviate target-domain sparsity by leveraging auxiliary-domain information \cite{ZangZLZY23}. In this context, compared with conventional CDR methods that mainly rely on behavioral similarity, KG-enhanced methods enable semantics-aware knowledge transfer, supporting higher-level reasoning and more flexible cross-domain association. Accordingly, integrating KGs into CDR has emerged as a promising direction and attracted growing research interest. Early studies, such as Fernández-Tobías et al.~\cite{Fernandez-Tobias19}, incorporate KG-derived semantic relatedness as auxiliary regularization in matrix factorization to establish inter-domain correlations. Subsequent works deepen KG utilization by focusing on either semantic or structural aspects, e.g., leveraging attribute-level semantics \cite{WangXTBLL25} or modeling structural dependencies through graph attention mechanisms \cite{Li0L23}. More recent approaches further combine semantic and structural modeling to capture complex entity correlations and enhance cross-domain transferability \cite{WangXTLYL24, LiHL24}.
	
\begin{figure}[]
		\centering
		\resizebox{0.65\linewidth}{!}{ 
			\includegraphics[width=\linewidth]{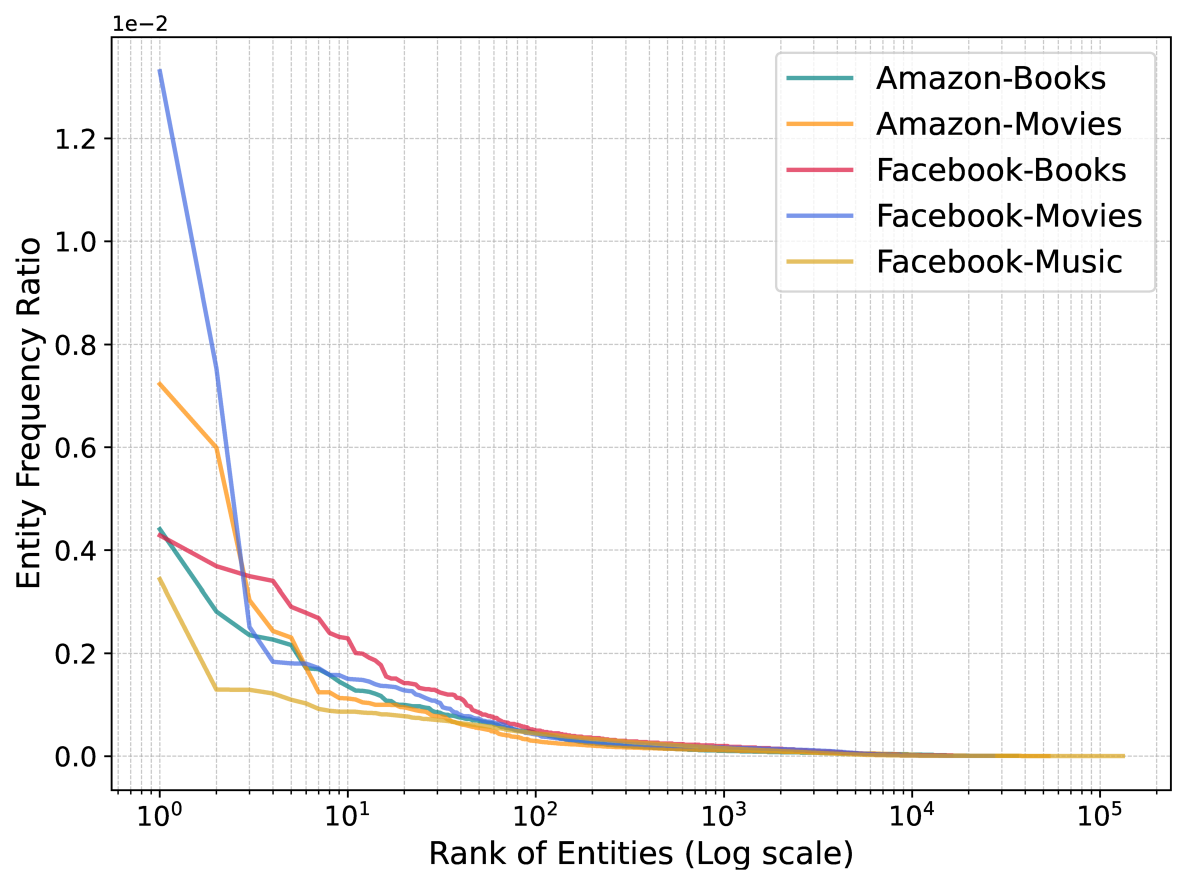}
		}
		\caption{Entity frequency distribution in real-world KGs}
		\label{fig:longtail}
	\end{figure}

	Despite these advances, KG-enhanced CDR methods still face practical challenges arising from KG sparsity, structural imbalance, and difficulties in domain alignment and semantic heterogeneity. We summarize these challenges as follows:
	
	\noindent \textbf{C1. Sensitivity to KG sparsity and popularity bias}. 
	Most KG-enhanced methods rely on static propagation paths or pre-trained knowledge graph embeddings (KGEs)~\cite{GuoZQZXXH22}, making them vulnerable to the structural imbalance of real-world KGs. In large-scale KGs, entity degrees typically follow a long-tail distribution~\cite{YuJ23, LiF17}, where a small fraction of popular entities dominate connectivity while the majority remain sparsely connected. As KGEs inherently encode graph structure, they tend to overemphasize high-degree entities and inadequately represent rare yet informative ones~\cite{wang2025large}. Consequently, downstream recommenders inherit popularity bias and exhibit limited generalization to cold-start or long-tail items.  We empirically verify this phenomenon on large-scale cross-domain KGs from Amazon~\cite{WangXTLYLData24} and Facebook~\cite{Fernandez-Tobias19}, where entity frequency statistics (Fig.~\ref{fig:longtail}) reveal pronounced long-tail patterns. We argue that in cross-domain settings, such entity-centric modeling further amplifies semantic imbalance across domains, as domain-specific knowledge is often concentrated in low-frequency entities, ultimately weakening cross-domain alignment and reducing the transferability of learned representations.
	
	\noindent \textbf{C2. Reliance on user overlap for cross-domain alignment}. Although KGs provide relational semantics that are potentially transferable across domains, most existing methods still depend on shared users as the primary anchors for cross-domain alignment, assuming the availability of overlapping users to establish behavioral correspondence. However, in real-world environments where platforms operate in isolation and user data sharing is constrained by privacy-preserving regulations, such as GDPR~\cite{gdpr2016} in the EU, CCPA~\cite{ccpa2018} in the USA, and PIPL~\cite{pipel2021} in China, user overlap is often unavailable \cite{YuanQCT0Y24}. In the absence of explicit behavioral anchors, conventional alignment strategies become ineffective, leading to severe performance degradation in non-overlapping scenarios. To bypass this reliance, recent works have explored Large Language Models (LLMs) to bridge domains semantically without shared users~\cite{00020W0Z0LH025, HadadRESR25, WangZL25, GuoLYNY24}. Nevertheless, the generative nature of LLMs inherently introduces substantial computational overhead and the risk of hallucinating factually ungrounded associations~\cite{LinDXLCZLWLZGYTZ25}. This vulnerability becomes particularly critical in non-overlapping scenarios, where the absence of shared users removes the necessary collaborative constraints to regularize the generative space.
	
	\noindent \textbf{C3. Lack of semantic disentanglement}. 
	Many CDR models implicitly assume that user preferences remain consistent across domains, aligning latent features without distinguishing between domain-shared and domain-specific semantics. In practice, \textit{domain-shared} semantics capture users’ stable and transferable interests, whereas \textit{domain-specific} semantics reflect contextual preferences unique to each domain  (e.g., niche film interests versus mainstream music tastes). Enforcing indiscriminate alignment across such heterogeneous signals can distort representation learning, leading to negative transfer and limited interpretability~\cite{WangYH2025, ZhangCLYP24}. Therefore, explicitly disentangling transferable and domain-specific semantics is critical for robust cross-domain transfer.
	
	\textbf{Design Motivation.} 
	To address the above challenges, we shift the modeling focus from entity-level representations to relation-level semantics in KGs. Compared with entities, relations are much fewer in number yet exhibit stronger semantic consistency, as each relation type connects a broad range of entities and forms denser structural patterns that are less sensitive to popularity bias. Moreover, relations capture high-level preference semantics by abstracting recurring patterns in user interests, making them well-suited to non-overlapping scenarios where direct structural alignment is unavailable.
	To validate the feasibility of relation-guided transfer, we analyze relation distributions in the aforementioned cross-domain KGs. As shown in Table~\ref{tab:relation-statistics}, several relations consistently appear across domains (e.g., \texttt{subject}, \texttt{basedOn}), while others are domain-specific (e.g., \texttt{televisionSeries} in movies, \texttt{stylisticOrigin} in books). This observation suggests that relation semantics simultaneously exhibit cross-domain consistency and domain discrimination, which motivates our knowledge-guided framework that encodes relations as \emph{soft prompts}~\cite{LiuYFJHN23}. These prompts provide structured semantic priors that dynamically guide preference modeling to facilitate effective cross-domain knowledge transfer.
	
		\begin{table*}[]
				\centering
				\caption{Knowledge Graph Relation Statistics across Domains. Domain-shared relations co-occur across domains, whereas domain-specific relations are confined to a single domain.}
				\label{tab:relation-statistics}
				\resizebox{0.95\linewidth}{!}{ 
						\begin{tabular}{@{}ccc cc@{}}
								\toprule
								\multirow{2}{*}{\textbf{Domain   Pair}} & \multicolumn{2}{c}{\textbf{Relation Num.}}        & \multicolumn{2}{c}{\textbf{Example Relation}}                                                                                                                                                                 \\ \cmidrule(lr){2-3} 
								\cmidrule(lr){4-5} 
								& \textbf{shared} & \textbf{specific} & \textbf{shared}                                                                                     & \textbf{specific}                                                                         \\ \midrule
								FB-Movie   \& Book                      & 25                     & \begin{tabular}[c]{@{}c@{}}Movie: 6\\ Book: 1\end{tabular}       & \begin{tabular}[c]{@{}c@{}}\texttt{subject}, \texttt{genre}, \texttt{literaryGenre} \\ \texttt{previousWork}, \texttt{basedOn} \end{tabular} & \begin{tabular}[c]{@{}c@{}}\texttt{televisionSeries}, \texttt{presenter}\\ \texttt{stylisticOrigin}\end{tabular}          \\ \midrule
								FB-Movie   \& Music                     & 28                     & \begin{tabular}[c]{@{}c@{}}Movie: 3 \\ Music: 2\end{tabular}       & \begin{tabular}[c]{@{}c@{}}\texttt{subject}, \texttt{previousWork}, \texttt{writer}, \\  \texttt{creator}, \texttt{associatedBand}\end{tabular}   & \begin{tabular}[c]{@{}c@{}} \texttt{televisionSeries}, \texttt{literaryGenre}\\ \texttt{movement}, \texttt{endingTheme}\end{tabular}       \\ \midrule
								FB-Music   \& Book                      & 24                     & \begin{tabular}[c]{@{}c@{}}Music: 6 \\ Book: 2\end{tabular}        & \begin{tabular}[c]{@{}c@{}}\texttt{subject}, \texttt{writer}, \texttt{genre}, \\ \texttt{award}, \texttt{basedOn}, \texttt{author}\end{tabular}                  & \begin{tabular}[c]{@{}c@{}}\texttt{movement}, \texttt{musicalBand}\\ \texttt{literaryGenre}, \texttt{stylisticOrigin}\end{tabular} \\ \midrule
								AM-Movie   \& Book                      & 17                     & \begin{tabular}[c]{@{}c@{}}Movie: 3\\ Book: 3\end{tabular}        & \begin{tabular}[c]{@{}c@{}}\texttt{series}, \texttt{award}, \texttt{genre},  \\ \texttt{subject}, \texttt{literaryGenre} \end{tabular}                    & \begin{tabular}[c]{@{}c@{}}\texttt{openingTheme}, \texttt{basedOn} \\ \texttt{notableWork}, \texttt{portrayer}\end{tabular}                   \\ \bottomrule
							\end{tabular}
				}\end{table*}

	\textbf{Our Approach.}  Building on the above motivation, we propose \textit{KGBridge}, a knowledge-guided prompt learning framework for cross-domain sequential recommendation under non-overlapping settings. KGBridge comprises two core components: (1) \textit{KG-enhanced Prompt Encoder} that encodes relation-level semantics from KGs into soft prompts, providing structured and transferable priors for sequential preference modeling, and (2)\textit{Two-stage Training Paradigm} that integrates cross-domain pretraining with privacy-preserving fine-tuning to support effective transfer without user overlap. Specifically, 
	\textbf{For C1}, relation-level prompts guide sequential modeling, thereby reducing sensitivity to KG sparsity and popularity bias while preserving semantically consistent patterns across domains.  
	\textbf{For C2}, the two-stage paradigm enables user-independent knowledge transfer, where cross-domain pretraining captures transferable relational priors, and fine-tuning adapts domain-specific features without relying on overlapping users. \textbf{For C3}, KGBridge introduces a correspondence-driven disentanglement mechanism with contrastive regularization to explicitly separate domain-shared and domain-specific knowledge, ensuring semantic complementarity and stabilizing adaptation during fine-tuning.
	
	\textbf{Summary of Contributions.} 
	We revisit the role of knowledge graphs in cross-domain recommendation, particularly in non-overlapping user settings where KGs serve as the primary semantic bridge across domains. Motivated by an empirical analysis of KG structural properties, we propose \textit{KGBridge}, a knowledge-guided prompt learning framework that shifts from entity-centric embeddings to relation-level semantic modeling for robust and privacy-preserving knowledge transfer. Extensive experiments on benchmark datasets demonstrate consistent performance improvements and strong robustness under varying levels of KG sparsity.

	\section{Preliminaries and Definitions}
	We study cross-domain sequential recommendation under a practical \textit{non-overlapping user} setting, where user identities are fully isolated across domains due to privacy and platform constraints. For each domain $d\in\{s,t\}$, let $\mathcal{U}_d$ and $\mathcal{V}_d$ denote the user and item sets, respectively. 
	The sequential recommendation data is denoted as $\mathcal{D}_d = \{(\mathcal{S}_u, y_u)\}_{u \in \mathcal{U}_d}$, 
	where $\mathcal{S}_u = [v_1, v_2, \ldots, v_N]$ represents the user’s historical interaction sequence, 
	and $y_u \in \mathcal{V}_d$ denotes the ground-truth next item following the sequence $\mathcal{S}_u$. 
	We focus on the non-overlapping scenario where $\mathcal{U}_s\cap\mathcal{U}_t=\emptyset$ and $\mathcal{V}_s\cap\mathcal{V}_t=\emptyset$, 
	which precludes direct user-level alignment and makes cross-domain knowledge transfer particularly challenging.
	
	Let $\mathcal{G}=(\mathcal{E},\mathcal{R},\mathcal{T})$ denote a knowledge graph (KG), where $\mathcal{E}$ and $\mathcal{R}$ represent entity and relation sets. $\mathcal{T}=\{(h,r,o)\}$ is the set of factual triples encoding relational semantics, where $h$ and $o$ denote the head and tail entities, and $r$ represents the relation type. For example, triples such as ``\emph{Titanic}, genre, \emph{Romance}'' or ``\emph{Inception}, actor, \emph{Leonardo DiCaprio}'' capture associations among entities. Such structured relational knowledge provides rich semantic priors that can benefit user preference modeling.
	
	\textbf{Problem Definition.} 
	Given source-domain data $\mathcal{D}_s$, target-domain data $\mathcal{D}_t$, and KG $\mathcal{G}$, 
	our goal is to learn a recommendation function
	$f_\Theta: (\mathcal{S}_u^t, \mathcal{G}) \rightarrow \hat{y}_u^t$, 
	where $\hat{y}_u^t$ denotes a predicted probability distribution over candidate items in the target domain. 
	The model is optimized by minimizing the prediction loss defined over the predicted probability assigned to the ground-truth item $y_u^t$.
	The basic notations are summarized in Table~\ref{tab:notations}.

	\begin{table}[]
		\centering
		\caption{Notation Summary. For clarity, domain superscripts (i.e., $s$ for source, $t$ for target) are omitted for general symbols unless explicit domain distinction is required.}
		\label{tab:notations}
		\begin{tabular}{l | l}
			\toprule
			\textbf{Symbol} & \textbf{Meaning} \\
			\midrule
			$\mathcal{G}$ & a knowledge graph with entities, relations, and triples \\
			$\mathcal{E}, \mathcal{R}, \mathcal{T}$ & sets of entities, relation types, and factual triples \\
			$(h, r, o)$ & a triple representing a semantic fact in the KG \\
			$\mathbf{R}_{\ast}$ & relation embeddings learned from the KG model, $\ast\in\{\text{shared}, \text{spec}\}$ \\
			$\mathbf{P}_{\ast}$ & KG-enhanced prompt banks derived from $\mathbf{R}_{\ast}$\\ \midrule
			$\mathcal{D}_d $ & sequential recommendation data in domain $d \in \{s, t\}$ \\
			$\mathcal{U}_s, \mathcal{U}_t$ & user sets in the source and target domains \\
			$\mathcal{V}_s, \mathcal{V}_t$ & item sets in the source and target domains \\
			$\mathcal{S}_u $ & historical interaction sequence of user $u$ \\
			$\mathbf{E}_u$ & initial sequence representation \\
			$\hat{\mathbf{e}}_i$ & knowledge-enriched item representation \\
			$\mathbf{z}_u$ & latent representation encoding dynamic user preferences \\ \midrule
			$y_u, \hat{y}_u$ & ground-truth and predicted next items for user $u$ \\
			$f_\Theta(\cdot)$ & recommendation function parameterized by $\Theta$ \\
			\bottomrule
		\end{tabular}
	\end{table}
	
	\section{Methodology}
	Building on the design motivation introduced above, this section presents the proposed \textit{KGBridge} framework in detail. We first provide an overview of the framework, then elaborate on the two core components of KGBridge: (1) \textbf{KG-enhanced Prompt Encoder}, which encodes relation-level semantics from knowledge graphs into semantic prompts to capture transferable and domain-sensitive semantics, guiding the subsequent sequence modeling process, and (2) \textbf{Two-stage Training Paradigm}, which enables effective and privacy-preserving cross-domain knowledge transfer under non-overlapping user settings.
	
	\subsection{Framework Overview}
\begin{figure*}[]
		\centering
		\resizebox{1\linewidth}{!}{ 
			\includegraphics[width=\linewidth]{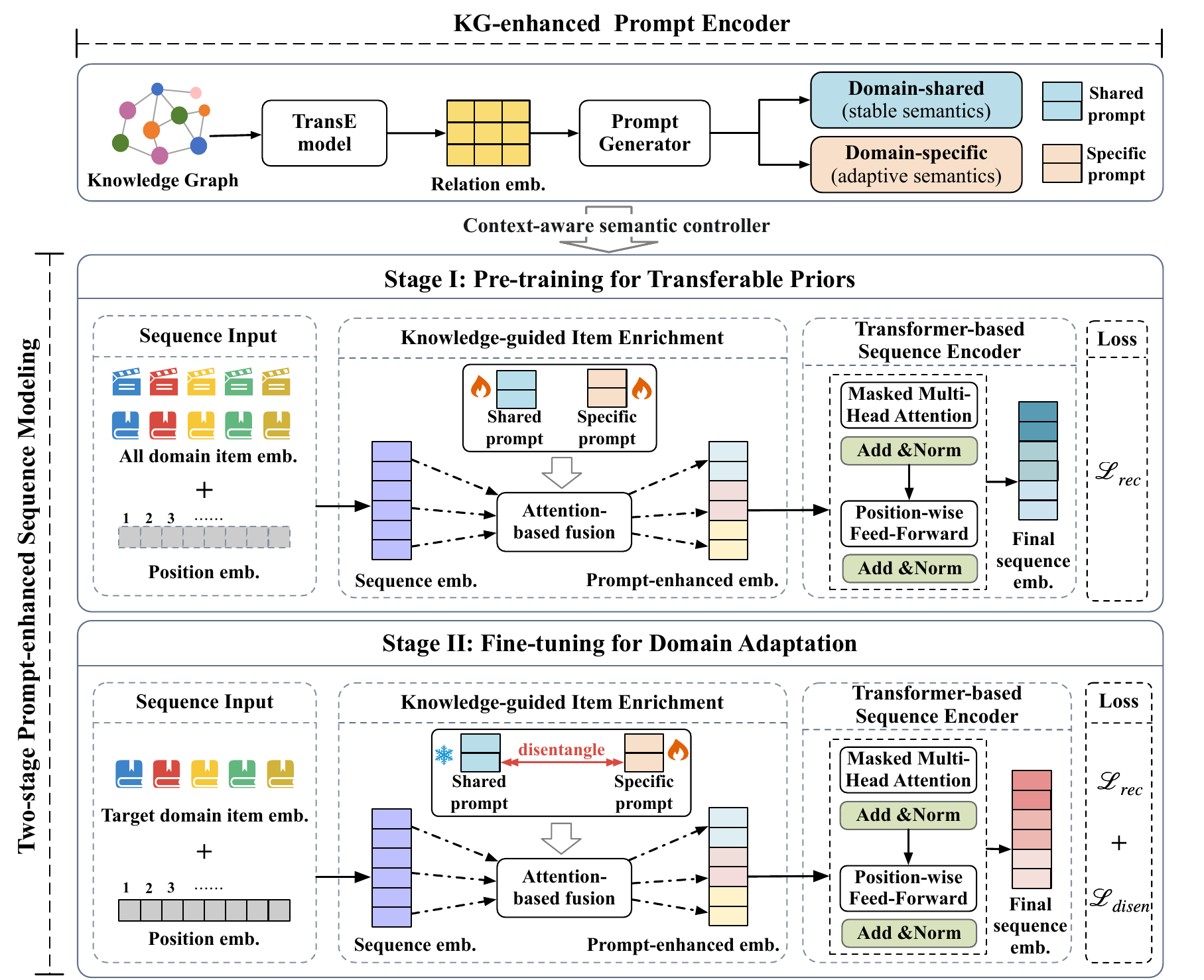}
		}
		\caption{Overview of the proposed KGBridge}
		\label{fig:framework}
	\end{figure*}

	As illustrated in Fig.~\ref{fig:framework}, \textit{KGBridge} centers on a \textit{KG-enhanced prompt encoder} and adopts a \textit{two-stage training paradigm}. The prompt encoder models relation-level semantics from KG into adaptive soft prompts, which are injected into the sequential recommendation model to guide domain-sensitive preference modeling. To enable effective transfer under non-overlapping user constraints, KGBridge adopts a two-stage training paradigm consisting of: (1) \textit{Pretraining}, which captures transferable relational priors from the source domain using KG-guided prompts, and (2) \textit{Fine-tuning}, which adapts these priors to the target domain without requiring direct cross-domain alignment. Through the joint design of relation-guided prompt injection and stage-wise optimization, KGBridge achieves effective cross-domain knowledge transfer while remaining robust to KG sparsity and the absence of explicit user-level correspondence.

	\subsection{KG-enhanced Prompt Encoder}
	The proposed framework is grounded in a \textit{KG-enhanced Prompt Encoder} (illustrated at the top of Fig.~\ref{fig:framework}), which integrates structured semantics from knowledge graphs into learnable \textit{soft prompts} to condition the sequential model. Soft prompts are continuous, learnable embeddings that operate directly in the representation space of model, rather than relying on discrete natural language tokens \cite{LiuYFJHN23}. Unlike hard or textual prompts, soft prompts possess independent parameters optimized for downstream objectives, enabling more flexible and fine-grained control over model behavior. 
	
	In our setting, we prioritize relation-level semantics over entity-level modeling, which is prone to severe long-tail effects in KGs. Relations are considerably fewer in number, more stable across domains, and less influenced by popularity bias, thus providing a more compact and transferable semantic basis. Based on this design, the encoder constructs two prompt banks: a \textbf{domain-shared bank} that captures invariant semantics transferable across domains, and a \textbf{domain-specific bank} that preserves distinctive relational knowledge within each domain.
	
	\subsubsection{Relation embedding learning} 
	Let $\mathcal{R}^{(s)}$ and $\mathcal{R}^{(t)}$ denote the KG relation vocabularies related to the source and target domains, respectively. We learn continuous embeddings for these relations to encode structured semantics, which serve as the semantic basis for subsequent prompt construction. To facilitate semantic disentanglement across domains, we first partition the relation vocabulary based on relation occurrence patterns:
	\begin{equation}
		\mathcal{R}_{\text{shared}} = \mathcal{R}^{(s)} \cap \mathcal{R}^{(t)}, \quad
		\mathcal{R}_{\text{spec}} = \big( \mathcal{R}^{(s)} \cup \mathcal{R}^{(t)} \big) \setminus \mathcal{R}_{\text{shared}} .
	\end{equation}
	This structural partition enables explicit modeling of domain-shared and domain-specific relational semantics, laying the groundwork for disentangled prompt construction during cross-domain transfer.
	
	KGs encode factual semantic associations among entities. To embed these structured semantics from the KG, we employ TransE \cite{BordesUGWY13}, a widely adopted and computationally efficient knowledge graph embedding method. TransE enforces a translational principle in the embedding space, where $\mathbf{h} + \mathbf{r} \approx \mathbf{o}$, and optimizes a margin-based ranking loss during training:
	\begin{equation}
		\mathcal{L}_{\text{TransE}} = 
		\sum_{(h,r,o)\in \mathcal{G}} \sum_{(h',r,o') \in \mathcal{G}'} 
		\big[ \gamma + \lVert \mathbf{h} + \mathbf{r} - \mathbf{o} \rVert_2 - \lVert \mathbf{h}' + \mathbf{r} - \mathbf{o}' \rVert_2 \big]_+ ,
	\end{equation}
	where $\mathcal{G}$ is the set of observed triples, $\mathcal{G}'$ denotes negative samples, $\gamma$ is a margin hyperparameter, and $[\cdot]_+$ is the hinge function.  
	
	\subsubsection{Prompt initialization via relation aggregation}\label{sec:prompt_agg}
	Although relation embeddings are more stable than entity embeddings, directly using a variable number of relations as prompts is suboptimal due to cross-domain discrepancies in relation cardinality and distribution. To address this issue, we design two dedicated prompt generators for domain-shared and domain-specific relations, respectively. Each generator aggregates a variable-sized set of relation embeddings into a fixed-length prompt matrix, thereby standardizing semantic representations for downstream modeling:
	\begin{equation}
		\mathbf{P}_{\ast} = \mathrm{PromptGenerator}_{\ast}(\mathbf{R}_{\ast}) \in \mathbb{R}^{L \times d}, \ast \in \{\text{shared}, \text{spec}\},
	\end{equation}
	where $\mathbf{R}_{\text{shared}} \in \mathbb{R}^{|\mathcal{R}_{\text{shared}}| \times d}$ and $\mathbf{R}_{\text{spec}} \in \mathbb{R}^{|\mathcal{R}_{\text{spec}}| \times d}$ are the relation embedding matrices learned by TransE, and $d$ denotes the embedding dimension. Each generator outputs $L$ prompt vectors that function as semantic controllers in downstream modeling. 
	
\subsubsection{Implementation and Theoretical Justification of PromptGenerator} \label{sec:prompt_gen_impl}
For $\mathrm{PromptGenerator}(\cdot)$, the key challenge is to transform a variable number of relations into stable semantic prompts. Since the aggregation strategy directly affects the trade-off between semantic stability and expressive capacity, we investigate several alternatives, including \textit{mean} pooling, \textit{attention} weighting, \textit{transformer}-based modeling, and our proposed \textit{mean\_noise}. Detailed empirical comparisons are provided in Section~\ref{sec:agg_study}.

\noindent \textbf{1) Deterministic Aggregation Baselines.}
We first consider conventional deterministic aggregation strategies:
\begin{itemize}[leftmargin=*]
	\item \textit{Pooling-based Aggregation:}
	Relation embeddings are aggregated into a single vector $\mathbf{h}\in\mathbb{R}^d$ via mean or max pooling, followed by an Multi-Layer Perceptron (MLP) to generate the $L \times d$ prompt matrix. Mean pooling captures global semantic tendencies, while max pooling emphasizes salient relation features. However, such fixed aggregation may oversmooth relation semantics or discard informative structural cues.
	\item \textit{Attention and Transformer-based Aggregation:}
	To model dependencies among relations, we further explore adaptive attention weighting and transformer-based encoding. Although more expressive, these approaches tend to overemphasize dominant or frequently co-occurring relations, causing the learned prompts to focus excessively on generic semantic patterns while weakening sparse yet discriminative signals. Moreover, the increased model complexity may reduce robustness under sparse cross-domain supervision.
\end{itemize}

\noindent \textbf{2) Mean Pooling with Stochastic Perturbation.}
To balance semantic stability and representation diversity, we adopt a lightweight stochastic aggregation strategy. Specifically, mean-aggregated relation embeddings are first projected into a semantic prototype $\bar{\mathbf{p}}\in\mathbb{R}^d$ through a linear layer:

\begin{equation}
	\bar{\mathbf{p}} = \mathbf{W}_{p} \left( \frac{1}{|\mathcal{R}_{\ast}|} \sum_{r \in \mathcal{R}_{\ast}} \mathbf{r} \right) + \mathbf{b}_{p}.
\end{equation}

We then replicate this prototype $L$ times and inject independent Gaussian perturbations to construct the prompt matrix:
\begin{equation}
	\mathbf{P}_{\text{ours}} =
	\big[
	\bar{\mathbf{p}} + \epsilon_1,\;
	\bar{\mathbf{p}} + \epsilon_2,\;
	\dots,\;
	\bar{\mathbf{p}} + \epsilon_L
	\big]^\top,
	\quad
	\epsilon_l \sim \mathcal{N}(0,\sigma^2\mathbf{I}).
\end{equation}

This stochastic perturbation mechanism (denoted as \textit{mean\_noise} in our experiments, with $\sigma=0.1$) serves as an implicit regularizer similar to the reparameterization strategy in the Variational Information Bottleneck framework~\cite{alemi2016deep}. Instead of fitting exact relation combinations, the injected noise encourages the encoder to capture stable semantic patterns shared across domains while preserving prompt diversity and improving robustness to sparse or imbalanced KG structures.

	\subsection{Two-stage Training Paradigm}
	Building on the semantic foundation established by the KG-enhanced Prompt Encoder, our framework adopts a two-stage learning strategy inspired by the paradigm of transferable pretraining and domain-specific adaptation \cite{LiuGZJGY24}. Distinct from previous works, we incorporate relation-guided prompts derived from KGs to inject structured relational semantics into the training process. These prompts provide explicit semantic guidance that facilitates cross-domain transfer and enhances model generalization, particularly under non-overlapping scenarios where direct user correspondence is unavailable. Specifically, the first stage jointly learns transferable priors shared across all domains, while the second stage refines the model to capture domain-specific characteristics under privacy-preserving constraints.
	
	\subsubsection{Stage I: Pretraining for Transferable Priors}
	In the pretraining stage (middle of Fig.~\ref{fig:framework}), the model is trained jointly on all available domains to learn domain-invariant relational patterns. Relation-guided prompts are injected as semantic controllers to augment item representations through an attention-based fusion mechanism, after which a Transformer encoder models temporal dependencies in user interaction sequences. This joint training process enables the model to acquire generalized relational patterns and stable semantic priors, forming a transferable representation space for subsequent adaptation in the fine-tuning stage.
	
	\paragraph{Sequence Input Representation}
	Given a user interaction sequence $\mathcal{S}_u = [v_1, v_2, \ldots, v_N]$, where $v_i$ denotes the $i$-th interacted item, we first construct its sequential representation by combining item and positional embeddings, following transformer-inspired designs \cite{VaswaniSPUJGKP17, KangM18, ZhangZLSXWLZ19}. This design allows the model to encode the relative positional dependencies among items within each user sequence:
	\begin{equation}
		\mathbf{E}_u = \mathrm{ItemEmb}(v_1, \ldots, v_N) + \mathrm{PosEmb}(1,\ldots,N),
		\label{eq:init_encode}
	\end{equation}
	where $\mathrm{ItemEmb}(\cdot)$ generates initial item embeddings, and $\mathrm{PosEmb}(\cdot)$ injects sequential information. The resulting sequence embedding $\mathbf{E}_u$ serves as the base representation for subsequent knowledge-guided enrichment and temporal modeling, with dropout applied to prevent overfitting.
	
	\paragraph{Knowledge-guided Item Enrichment}
	The sequence representation $\mathbf{E}_u$ captures the ordered behavioral context of each user, while relation-guided prompts provide complementary KG semantics. To integrate these signals, we design a Knowledge-guided Item Enrichment module that conditions position-aware item embeddings on the prompts before sequential encoding. For each user sequence, we obtain position-aware item embeddings $\mathbf{E}_{u}\in\mathbb{R}^{N\times d}$, and two prompt banks $\mathbf{P}_{\mathrm{shared}}\in\mathbb{R}^{L\times d}$ and $\mathbf{P}_{\mathrm{spec}}\in\mathbb{R}^{L\times d}$ generated by the KG-enhanced Prompt Encoder.
	To provide consistent relational guidance across temporal positions, the two prompt banks are shared across the sequence and jointly contextualize each position-aware item embedding. Specifically, the prompt--item context at position $i$ is defined as
	

	\begin{equation}
		\mathbf{X}_{u,i}
		=
		[\mathbf{P}_{\mathrm{shared}};
		\mathbf{P}_{\mathrm{spec}};
		\mathbf{e}_{i}]
		\in\mathbb{R}^{(2L+1)\times d},
		\qquad i=1,\ldots,N,
	\end{equation}
	where $\mathbf{e}_{i}$ denotes the $i$-th row of $\mathbf{E}_{u}$. Applying this construction to all positions yields
	$\mathbf{X}_{u}\in\mathbb{R}^{N\times(2L+1)\times d}$,
	which corresponds to broadcasting the prompt banks along the sequence dimension. In this way, the prompts provide a common relational basis throughout the sequence, while the subsequent attention mechanism determines their item-dependent contributions.
	
	To enable adaptive prompt influence rather than treating prompts as static context, we design an attention-based fusion mechanism over $\mathbf{X}_u$. Specifically, for each sequence position $i$, we model the interaction between each prompt and the corresponding item by concatenating their embeddings, forming $\mathbf{h}_{i,j} = [\mathbf{X}_{u,i,j}; \mathbf{e}_i] \in \mathbb{R}^{2d}$, where $\mathbf{e}_{i}$ denotes the $i$-th row of sequence embedding $\mathbf{E}_u$ defined in Eq. (\ref{eq:init_encode}). This joint representation encodes the compatibility between the $j$-th prompt and the $i$-th item. A feed-forward attention network $f_{\text{att}}(\cdot)$ is then applied to estimate the relative contribution of each prompt:
	\begin{equation}
		\alpha_{i,j} = \frac{\exp(f_{\text{att}}(\mathbf{h}_{i,j}))}{\sum_{k} \exp(f_{\text{att}}(\mathbf{h}_{i,k}) )}.
	\end{equation}
	Finally, the knowledge-enriched item representation is obtained as a weighted combination:
	\begin{equation}
		\hat{\mathbf{e}}_i = \sum_{j} \alpha_{i,j} \mathbf{X}_{u,i,j}.
	\end{equation}
	
	Through this design, prompts serve as context-aware semantic controllers, dynamically adjusting item representations based on both relational relevance and sequence context. Specifically, domain-shared prompts emphasize invariant semantics beneficial for cross-domain transfer, while domain-specific prompts refine unique relational nuances in each domain. The resulting enriched embeddings integrate KG semantics with sequential dependencies, providing informative inputs for the subsequent sequence encoder.
	
	\paragraph{Sequence Encoder}
	The prompt-enriched item embeddings $\{\hat{\mathbf{e}}_1,\ldots,\hat{\mathbf{e}}_N\}$ are subsequently fed into a Transformer-based sequence encoder to capture user preference dynamics. Following the paradigm of self-attentive sequential recommenders \cite{KangM18, SunLWPLOJ19, ZhangZLSXWLZ19}, the encoder employs multi-head self-attention to model both short-term transitions and long-range dependencies within user interaction sequences. Within each layer, the encoder computes Query–Key–Value attention among all items under a causal mask, restricting attention to historical items to preserve temporal consistency. A position-wise feed-forward network further transforms the representations, while residual connections and layer normalization after every sublayer stabilize training and facilitate gradient propagation. Formally, the final sequence representation for user $u$ is computed as:
	\begin{equation}
		\mathbf{z}_u = \mathrm{TransformerEncoder}([\hat{\mathbf{e}}_1, \ldots, \hat{\mathbf{e}}_N]),
	\end{equation}
	where $\mathbf{z}_u \in \mathbb{R}^d$ encodes the dynamic user preferences along the sequence. Since this component is not the core contribution of our framework and follows well-established sequential modeling practices, we briefly summarize its implementation here for clarity.

	\paragraph{Pretraining Objective}
	The pretraining stage is formulated as a next-item prediction task, where the model predicts the subsequent item based on a user's historical interaction sequence. Specifically, given a user's historical interactions $\mathcal{S}_{u,\leq t}$, the model outputs a hidden sequence representation $\mathbf{z}_u$, which is mapped into item-wise logits via a linear classifier: $\mathbf{s}_u = \mathbf{W}\cdot\mathbf{z}_u + \mathbf{b}$, where $\mathbf{W} \in \mathbb{R}^{|\mathcal{V}|\times d}$ and $\mathbf{b} \in \mathbb{R}^{|\mathcal{V}|}$ are learnable parameters, and $|\mathcal{V}|$ denotes the number of candidate items.  The probability of recommending item $i$ is obtained through softmax normalization:
	\begin{equation}
		p(i|\mathcal{S}_{u,\leq t}) = \frac{\exp(s_{u,i})}{\sum_{j \in \mathcal{V}} \exp(s_{u,j})}.
		\label{eq:prediction}
	\end{equation}
	
	The model is trained using a cross-entropy loss over observed interactions from all domains:
	\begin{equation}
		\mathcal{L}_{\text{pretrain}} = - \sum_{u \in \mathcal{U}} \sum_{t=1}^{T} \log p(v_{t+1}|\mathcal{S}_{u,\leq t}), 
	\end{equation}
	which encourages higher likelihoods for ground-truth items while suppressing others. During this stage, parameters of both the domain-shared and domain-specific prompts, as well as the sequence encoder, are jointly optimized over all domains. This joint pretraining enables the model to internalize cross-domain prior knowledge, providing semantic-aware initialization for privacy-preserving fine-tuning in the target domain.
	
	\subsubsection{Stage II: Fine-tuning for Domain Adaptation}
	After the pretraining stage, the model is fine-tuned on the target domain (bottom of Fig.~\ref{fig:framework}) to capture domain-specific features while preserving cross-domain semantic consistency. This stage is designed for non-overlapping settings, enabling privacy-preserving adaptation without requiring shared users or items.
	
	\paragraph{Parameter Freezing and Adaptive Update}
	To balance generalization and specialization, parameters encoding domain-invariant knowledge are frozen, while domain-adaptive components are further optimized. Specifically, the domain-shared prompts $\mathbf{P}_{\text{shared}}$, which encapsulate transferable relational invariances, are fixed during fine-tuning and act as semantic anchors that stabilize adaptation. Meanwhile, the domain-specific prompts $\mathbf{P}_{\text{spec}}$ and the sequence encoder are updated to capture target-domain characteristics that complement the shared knowledge. This selective adaptation allows the model to adapt effectively to the target domain while retaining the relational priors established during pretraining.
	
	\paragraph{Prompt Disentanglement}
	While parameter freezing stabilizes the domain-shared prompts by preventing their update, it alone cannot regulate how the shared and specific representations interact during target-domain adaptation. Recent studies \cite{WangXBTLL25, LocatelloBLRGSB19} highlight the importance of explicit supervision for organizing complementary representations in disentangled learning. However, existing disentangled CDR methods \cite{LiuGZJGY24, GuoWWZY25} predominantly rely on unsupervised constraints, such as orthogonality or alternating optimization, which provide limited guidance for coordinating shared and specific representations.
	
	To address this limitation, we design a \textit{correspondence-driven disentanglement} strategy, which provides a soft yet explicit regularization signal to guide the prompt interaction in a controlled manner. 
	
	Given the shared and specific prompt banks
	$\mathbf{P}_{\mathrm{shared}}=\{\mathbf{p}_{i}^{(\mathrm{sh})}\}_{i=1}^{L}$
	and
	$\mathbf{P}_{\mathrm{spec}}=\{\mathbf{p}_{i}^{(\mathrm{sp})}\}_{i=1}^{L}$,
	the two PromptGenerators independently produce $L$ latent prompt vectors, without assuming any semantic correspondence between same-index vectors before optimization. During fine-tuning, we impose a positional correspondence by treating
	$(\mathbf{p}_{i}^{(\mathrm{sh})},\mathbf{p}_{i}^{(\mathrm{sp})})$
	as a positive same-index pair and
	$(\mathbf{p}_{i}^{(\mathrm{sh})},\mathbf{p}_{j}^{(\mathrm{sp})})$, where $j\neq i$,
	as cross-index negative pairs. Based on this correspondence, we apply the following InfoNCE-based objective:
	
	\begin{equation}
		\mathcal{L}_{\text{disen}} = 
		-\frac{1}{L} \sum_{i=1}^{L} 
		\log 
		\frac{\exp\left(\mathrm{sim}(\mathbf{p}^{(\text{sh})}_i, \mathbf{p}^{(\text{sp})}_i)/\tau\right)}
		{\sum_{j=1}^{L} \exp\left(\mathrm{sim}(\mathbf{p}^{(\text{sh})}_i, \mathbf{p}^{(\text{sp})}_j)/\tau\right)},
	\end{equation}
	where $\operatorname{sim}(\cdot,\cdot)$ denotes cosine similarity and $\tau$ is the temperature parameter. This objective encourages each specific prompt to maintain stronger relative correspondence with its same-index shared prompt than with cross-index prompt slots. Together with the frozen shared prompts and the target-domain recommendation objective, this correspondence provides structured guidance for adapting the specific prompts around stable shared relational references.
	
	\noindent \textbf{Design Motivation.}
	The design reflects a structured dependency between prompt types: shared prompts encode transferable relational regularities, whereas specific prompts capture domain-sensitive refinements. During fine-tuning, the shared prompts remain fixed as stable references, while the specific prompts are adaptively updated under target-domain supervision. The correspondence regularization further coordinates the two prompt banks by encouraging each specific prompt to remain relatively closer to its same-index shared reference than to other prompt slots, thereby providing structured guidance for target-domain adaptation.
	
	\paragraph{Fine-tuning Objective and Prediction}
	During fine-tuning, next-item prediction follows the same formulation as in pretraining, where the probability of each candidate item is estimated via the softmax normalization in Eq.~(\ref{eq:prediction}), yielding the recommendation loss $\mathcal{L}_{\text{rec}}$. The overall fine-tuning objective jointly optimizes the recommendation loss with the disentanglement regularization:
	\begin{equation}
		\mathcal{L}_{\text{finetune}} = \mathcal{L}_{\text{rec}} + \lambda \mathcal{L}_{\text{disen}},
		\label{eq:finetune}
	\end{equation}
	where $\mathcal{L}_{\text{rec}}$ denotes the cross-entropy loss over target-domain interactions, and $\lambda$ controls the regularization strength. 
	
	During optimization, domain-shared prompts are kept frozen to preserve transferable relational priors, while domain-specific prompts are adaptively updated under correspondence-driven contrastive supervision. This combination of parameter freezing and disentangled regularization enables the model to capture domain-specific patterns without semantic drift, ensuring stable and effective adaptation in non-overlapping settings.
	
	\subsection{Complexity Analysis}
	To demonstrate the efficiency of KGBridge, we analyze its time and space complexity in comparison with standard sequential recommenders and KG-enhanced baselines.
	
	\paragraph{Time Complexity} 
	The computational cost of KGBridge mainly arises from three components: relation-guided prompt initialization, prompt-enhanced sequential encoding, and disentanglement regularization.
	
	For prompt initialization, aggregating relation embeddings requires $\mathcal{O}(|\mathcal{R}| \cdot d)$, where $|\mathcal{R}|$ denotes the number of relations and $d$ is the embedding dimensionality. Since the number of relations is typically much smaller than the number of entities (i.e., $|\mathcal{R}| \ll |\mathcal{E}|$), this overhead is negligible in practice.
	
	During sequential encoding, concatenating $2L$ soft prompts (including shared and domain-specific prompts) with an interaction sequence of length $N$ results in a self-attention complexity of $\mathcal{O}((N+2L)^2 \cdot d)$. Since $L$ and $N$ are bounded hyperparameters in practice (e.g., $L=2$, $N=15$), the additional prompts introduce only marginal overhead, preserving the same asymptotic order as standard self-attention-based models such as SASRec~\cite{KangM18}.
	
	In addition, the disentanglement regularization only involves lightweight pairwise similarity computation over prompts, resulting in negligible overhead. In contrast, KG-enhanced methods such as KSR~\cite{HuangZDWC18} and KMGCDR~\cite{WangXTBLL25} rely on Key-Value Memory Networks (KV-MN), whose memory reading and updating operations introduce additional overhead proportional to memory size during sequential modeling. By replacing entity-centric memory interactions with compact relation-guided prompts, KGBridge achieves substantially improved efficiency while retaining effective knowledge transfer capability.
	
	\paragraph{Space Complexity} 
 	The learnable parameters of KGBridge mainly consist of item embeddings, relation embeddings, soft prompts, and Transformer parameters, resulting in a space complexity of
	$\mathcal{O}(|\mathcal{V}|d + |\mathcal{R}|d + 2Ld + \text{Transformer})$,
	where $2L$ denotes the total number of learnable prompts (shared and domain-specific). Although the prompts are initialized from aggregated relation semantics, they remain trainable during both pretraining and fine-tuning, introducing only lightweight additional parameters due to their small size.
	
	Unlike entity-centric KG-enhanced recommenders that maintain large entity embedding matrices $\mathcal{O}(|\mathcal{E}|d)$ derived from KGEs or large external memory matrices, KGBridge relies on compact relation-level semantics and a small set of continuous prompts for knowledge transfer. Therefore, it achieves favorable parameter efficiency while maintaining a memory footprint comparable to standard sequential recommendation models.

	\section{Experiments} \label{experiments}
	To comprehensively evaluate the proposed KGBridge, we conduct experiments to answer the following questions: 
	\begin{description}[leftmargin=1.8em]
		\item[Q1] How does KGBridge compare with state-of-the-art methods in overall recommendation performance?
		\item[Q2] How do the key design choices of KGBridge contribute to cross-domain semantic transfer?
		\item[Q3] To what extent are shared relational semantics consistent across domains?
		\item[Q4] How do different aggregation mechanisms balance semantic stability and expressive capacity in relation-guided prompt construction?
		\item[Q5] How robust is KGBridge when the underlying knowledge graph becomes increasingly sparse or incomplete?
		\item[Q6] Does the proposed two-stage training strategy enable stable optimization and effective cross-domain transfer?
		\item[Q7] How does KGBridge separate and coordinate shared and domain-specific relational information during target-domain adaptation?
		\item[Q8] How sensitive is KGBridge to the disentanglement balancing coefficient?
	\end{description}

	\subsection{Datasets}
	We conduct experiments on two widely used KG-based cross-domain recommendation datasets, Facebook and Amazon, both constructed based on the well-known knowledge graph \textit{DBpedia}\footnote{https://www.dbpedia.org} and extensively adopted in prior surveys and representative methods~\cite{LiLWCZ21, AnelliNSFM21, WangXTBLL25, WangXTLYLData24}. These datasets include multi-domain user–item interactions with structured KG information, enabling the investigation of knowledge transfer across domains. Specifically, the Facebook dataset \cite{Fernandez-Tobias19} covers user interactions and item metadata from three domains: \textit{Movie}, \textit{Book}, and \textit{Music}; while the Amazon dataset \cite{WangXTLYL24} comprises user ratings and product information from the \textit{Movies\_and\_TV} and \textit{Books} domains, where ratings greater than 3 are treated as positive feedback. To ensure consistency and high-quality KG alignment, we retain only items linked to DBpedia entities, following standard practices in KG-based recommendation \cite{HuangZDWC18, WangZWZLXG18, WangXTBLL25}. The detailed statistics of the processed datasets are summarized in Table~\ref{tab:Dataset}, where ``FB-'' and ``AM-'' denote Facebook and Amazon domains, respectively.
	
	As previously shown in Fig.~\ref{fig:longtail} and Table~\ref{tab:relation-statistics}, both datasets exhibit pronounced long-tail distributions in entity frequency and diverse relational patterns, posing challenges for stable and balanced knowledge transfer. Following the common dense-to-sparse setting in CDR, we construct four transfer tasks based on domain sparsity levels. Three tasks consider within-platform transfer: \textbf{FB Movie$\rightarrow$Book}, \textbf{FB Music$\rightarrow$Book}, and \textbf{AM Movie$\rightarrow$Book}; the fourth, \textbf{FB Music$\rightarrow$AM Books}, considers cross-platform transfer from Facebook to Amazon. Across all tasks, knowledge from an information-rich source domain is leveraged to alleviate limited behavioral and semantic supervision in the sparse target domain.
	
	\begin{table}[]
		\caption{Statistics of the RS and KG dataset}
		\label{tab:Dataset}
		\centering
		\setlength{\tabcolsep}{4pt}
		\begin{tabular}{@{}c|rrrrr@{}}
				\toprule
				Domain       & FB-Movie  & FB-Book & FB-Music  & AM-Movie & AM-Book \\ \midrule
				Users        & 54,265    & 23,931  & 72,017    & 5,648     & 10,485   \\
				Items        & 4,843      & 4,160   & 6,182     & 4,779     & 8,318   \\
				Inter. & 1,390,661 & 213,045 & 2,493,080 & 67,066    & 115,572 \\
				Sparsity     & 99.47\%   & 99.79\% & 99.44\%   & 99.75\%  & 99.87\% \\ 
				Avg. Seq. Len. & 25.63 & 8.90& 34.62 & 11.87 & 11.02\\\midrule
				Relation     & 31        & 26      & 30        & 20       & 20      \\
				Entities     & 26,924    & 16,130  & 132,168   & 53,096    & 35,995   \\
				Triples      & 96,122    & 38,376  & 359,454   & 151,515   & 80,684   \\ \bottomrule
			\end{tabular}
	\end{table}

	\subsection{Experimental setup}
	\subsubsection{Evaluation protocol}
	To rigorously evaluate model performance under non-overlapping conditions, we ensure that no identity-level user correspondence is available between the source and target domains. For the three within-platform tasks, we simulate this setting by independently re-indexing user identities across domains; for the cross-platform task, the Facebook and Amazon datasets provide no cross-platform user correspondence. Together, these settings allow us to evaluate non-overlapping transfer under
	both controlled within-platform identity isolation and naturally separated cross-platform user populations. All cross-domain methods are evaluated under this same protocol, with methods originally relying on overlapping users adapted by disabling the corresponding identity-dependent operations while retaining their remaining applicable components, as described in Section~\ref{sec:competitors}.
	For each domain, we follow the widely adopted leave-one-out strategy in sequential recommendation \cite{KangM18, SunLWPLOJ19}. Specifically, for each user sequence, the last item is held out for testing, the second-to-last for validation, and all preceding interactions are used for training. This setup aligns with the next-item prediction paradigm and reflects natural temporal dynamics in user behavior. 
	
	We adopt a full-ranking evaluation over all candidate items instead of sampling-based protocols to avoid evaluation bias and ensure consistent comparisons~\cite{Zhang0Z024}. Model performance is evaluated using two widely adopted metrics\cite{GuoH2025, LiMSLYHXZM24}, Recall@$K$ and NDCG@$K$, which respectively measure top-$K$ retrieval accuracy and ranking quality. We report results for $K=\{3,5,10,20\}$ to provide a comprehensive evaluation. Higher values of Recall@$K$ and NDCG@$K$ indicate better recommendation performance.  The experiments yield statistical significance after five independent runs.
	
	\subsubsection{Competitors}\label{sec:competitors}
	We compare KGBridge with representative state-of-the-art methods from both single-domain and cross-domain recommendation. Baselines are selected to cover three methodological dimensions most relevant to our work: (i) \textit{KG-aware modeling}, (ii) \textit{sequential user modeling}, and (iii) \textit{non-overlapping cross-domain transfer}. Since no existing method simultaneously addresses all three aspects, we adopt a complementary set of baselines spanning  KG-based, sequential, KG-enhanced sequential, non-overlapping cross-domain, and KG-enhanced cross-domain methods. This comprehensive selection enables a fair and systematic evaluation of the effectiveness of each design component in KGBridge.
	
	\textit{Cross-domain methods. }
	(1) \textbf{KMGCDR}~\cite{WangXTBLL25} integrates a knowledge-enhanced memory network into a GNN-based framework and exploits shared-user preferences for cross-domain fusion; this shared-user fusion is inactive because no user correspondence is available.
	(2) \textbf{MCRPL}~\cite{LiuGZJGY24} targets non-overlapping cross-domain recommendation by pretraining domain-shared and domain-specific prompts for knowledge transfer, but it does not incorporate KG semantics.
	(3) \textbf{LLM4CDSR}~\cite{00020W0Z0LH025}  employs LLM-derived textual item representations and hierarchical user profiling for cross-domain sequential recommendation. We adapt its original overlapping-user formulation by disabling the pairwise cross-domain regularization that requires same-user interactions, while retaining its shared semantic and sequential modeling components.
	
	\textit{Single-domain methods. }
	These baselines operate solely within the target domain. We group them into two categories:
	(1) \textit{Sequential models}, including \textbf{SASRec}~\cite{KangM18}, \textbf{SRGNN}~\cite{WuT0WXT19}, and \textbf{GRU4Rec}~\cite{TanXL16}, which capture temporal dependencies and user behavior dynamics. 
	(2) \textit{KG-based models}: \textbf{GRU4RecKG}, an extension of GRU4Rec that concatenates item with KG entity embeddings to incorporate external semantics; \textbf{KGAT}~\cite{Wang00LC19}, which models higher-order connectivity through attentive graph propagation; and \textbf{KSR}~\cite{HuangZDWC18}, which integrates a sequential encoder with a KG-enriched memory network to jointly capture sequential and attribute-level preferences.
	
	\subsubsection{Implementation and hyperparameter setting}
	All baseline models are implemented in PyTorch, primarily based on the open-source library RecBole~\cite{zhao2021recbole}, while KMGCDR and MCRPL are reproduced from their official implementations. All experiments are conducted on a single NVIDIA GeForce RTX~4090 GPU (24~GB memory). 
	For knowledge graph embedding initialization, we employ TransE implemented in the toolkit \texttt{OpenKE} \cite{han2018openke} with an embedding dimension of 100. During TransE pre-training, we adopt Bernoulli negative sampling with a negative-to-positive ratio of $25{:}1$, where either the head or tail entity is corrupted to generate negative triples. 
	
	To ensure fair comparison, we standardize key hyperparameters across all methods: the embedding dimension is fixed to 100, the maximum sequence length to 15, the batch size to 128, and the learning rate to $1\times10^{-4}$. Early stopping with a patience of 10 epochs is applied to prevent overfitting, and the dropout rate is tuned within $[0.1, 0.5]$. For our proposed KGBridge,  the temperature $\tau$ in the disentanglement loss is fixed to 0.2; the prompt length $L=2$; and the balancing coefficient $\lambda$ for $\mathcal{L}_{\mathrm{disen}}$ is selected from $\{0, 0.001, 0.002, 0.003, 0.005, 0.01, 0.1\}$, yielding $0.002$ for FB Movie--Books and AM Movie--Books and $0.003$ for FB Music--Books (see Section~\ref{sec:hyperparam} for detailed analysis). The source code and datasets are publicly available at \url{https://github.com/WangYuhan-0520/KGBridge}.

	\subsection{Overall performance (Q1)}
	\begin{table*}[t]
		\caption{Overall performance comparison. The best results are highlighted in bold, and the second-best are underlined. “$\dagger$” marks KG-based methods. “*” indicates statistical significance over the strongest baseline (paired t-test, p-value\textless0.05). }
		\setlength{\tabcolsep}{3.5pt} 
		\label{tab:OverallComp}
		\resizebox{1\linewidth}{!}{ 
			\begin{tabular}{@{}clcccccccccccc@{}} 
				\toprule
				\multirow{3}{*}{\textbf{Dataset}}                                              & \multicolumn{1}{c}{\multirow{3}{*}{\textbf{Metric}}} & \multicolumn{6}{c}{\multirow{2}{*}{\textbf{Single-domain Methods}}} & \multirow{3}{*}{} & \multicolumn{4}{c}{\multirow{2}{*}{\textbf{Cross-domain Methods}}} & \multirow{3}{*}{\textbf{Imp.}} \\
				& \multicolumn{1}{c}{}                                 & \multicolumn{6}{c}{}                                                &                   & \multicolumn{4}{c}{}                                               &                                             \\ \cmidrule(lr){3-8} \cmidrule(lr){10-13}
				& \multicolumn{1}{c}{}                                 & \textit{SASRec}        & \textit{SRGNN}   & \textit{GRU4Rec}   & \textit{GRU4RecKG}$\dagger$  & \textit{KGAT}$\dagger$  & \textit{KSR}$\dagger$   &                   & \textit{KMGCDR}$\dagger$ & \textit{LLM4CDSR} & \textit{MCRPL}             & \textit{KGBridge}$\dagger$       &                                             \\ \midrule
				\multirow{8}{*}{\begin{tabular}[c]{@{}c@{}}FB-Movie\\ $\downarrow$ \\FB-Books\end{tabular}} & R@3                                             & \underline{.0940}   & .0860   & .0904     & .0871       & .0347  & .0674  &                   & .0010 & .0364 & .0908             & \textbf{.0977 }                 & 3.94\%                                      \\
				& R@5                                             & \underline{.1228}   & .1106   & .1187     & .1128       & .0520  & .0894  &                   & .0211 & .0520 & .1191             & \textbf{.1275 }                 & 3.83\%                                      \\
				& R@10                                            & \underline{.1698}   & .1558   & .1668     & .1614       & .0939  & .1301  &                   & .0605 & .0798 & .1627             & \textbf{.1741*}                 & 2.53\%                                      \\
				& R@20                                            & \underline{.2294}   & .2134   & .2288     & .2260       & .1521  & .1848  &                   & .1323 & .1213 & .2209             & \textbf{.2333*}                 & 1.70\%                                      \\
				& N@3                                               & \underline{.0748}   & .0683   & .0726     & .0686       & .0241  & .0528  &                   & .0066 & .0261 & .0723             & \textbf{.0792*}                 & 5.88\%                                      \\
				& N@5                                               & \underline{.0867}   & .0785   & .0842     & .0792       & .0312  & .0619  &                   & .0112 & .0326 & .0839             & \textbf{.0914*}                 & 5.42\%                                      \\
				& N@10                                              & \underline{.1018}   & .0931   & .0996     & .0948       & .0447  & .0750  &                   & .0237 & .0415 & .0979             & \textbf{.1064 }                 & 4.52\%                                      \\
				& N@20                                              & \underline{.1168}   & .1075   & .1152     & .1110       & .0593  & .0887  &                   & .0417 & .0519 & .1126             & \textbf{.1213 }                 & 3.85\%                                      \\ \midrule
				\multirow{8}{*}{\begin{tabular}[c]{@{}c@{}}FB-Music\\ $\downarrow$ \\FB-Books\end{tabular}} & R@3                                             & \underline{.0940}   & .0860   & .0904     & .0843       & .0315  & .0691  &                   & .0077 & .0074 & .0860             & \textbf{.0972*}                 & 3.40\%                                      \\
				& R@5                                             & \underline{.1228}   & .1106   & .1187     & .1104       & .0506  & .0910  &                   & .0187 & .0137 & .1152             & \textbf{.1273 }                 & 3.66\%                                      \\
				& R@10                                            & \underline{.1698}   & .1558   & .1668     & .1580       & .0914  & .1333  &                   & .0562 & .0390 & .1608             & \textbf{.1773 }                 & 4.42\%                                      \\
				& R@20                                            & \underline{.2294}   & .2134   & .2288     & .2207       & .1537  & .1877  &                   & .1291 & .0781 & .2193             & \textbf{.2397 }                 & 4.49\%                                      \\
				& N@3                                               & \underline{.0748}   & .0683   & .0726     & .0667       & .0219  & .0553  &                   & .0054 & .0054 & .0687             & \textbf{.0775*}                 & 3.61\%                                      \\
				& N@5                                               & \underline{.0867}   & .0785   & .0842     & .0774       & .0298  & .0642  &                   & .0098 & .0080 & .0808             & \textbf{.0898*}                 & 3.58\%                                      \\
				& N@10                                              & \underline{.1018}   & .0931   & .0996     & .0927       & .0429  & .0778  &                   & .0218 & .0160 & .0954             & \textbf{.1058 }                 & 3.93\%                                      \\
				& N@20                                              & \underline{.1168}   & .1075   & .1152     & .1085       & .0585  & .0915  &                   & .0401 & .0259 & .1101             & \textbf{.1215 }                 & 4.02\%                                      \\ \midrule
				\multirow{8}{*}{\begin{tabular}[c]{@{}c@{}}AM-Movie\\ $\downarrow$ \\AM-Books\end{tabular}} & R@3                                             & .0343         & .0234   & .0290     & .0325       & .0175  & .0223  &                   & .0068 & .0130 & \underline{.0382}       & \textbf{.0400*}                 & 4.71\%                                      \\
				& R@5                                             & .0495         & .0347   & .0443     & .0466       & .0281  & .0337  &                   & .0140 & .0197 & \underline{.0539}       & \textbf{.0608*}                 & 12.80\%                                     \\
				& R@10                                            & .0735         & .0561   & .0739     & .0742       & .0569  & .0546  &                   & .0332 & .0360 & \underline{.0822}       & \textbf{.0943*}                 & 14.72\%                                     \\
				& R@20                                            & .1090         & .0944   & .1171     & .1114       & .0993  & .0831  &                   & .0753 & .0577 & \underline{.1229}       & \textbf{.1380*}                 & 12.29\%                                     \\
				& N@3                                               & .0254         & .0177   & .0213     & .0247       & .0127  & .0166  &                   & .0047 & .0090 & \underline{.0294}       & \textbf{.0298 }                 & 1.36\%                                      \\
				& N@5                                               & .0316         & .0224   & .0276     & .0305       & .0170  & .0212  &                   & .0076 & .0118 & \underline{.0358}       & \textbf{.0383*}                 & 6.98\%                                      \\
				& N@10                                              & .0393         & .0292   & .0370     & .0394       & .0262  & .0278  &                   & .0137 & .0171 & \underline{.0450}       & \textbf{.0492*}                 & 9.33\%                                      \\
				& N@20                                              & .0483         & .0388   & .0478     & .0487       & .0368  & .0350  &                   & .0242 & .0225 & \underline{.0553}       & \textbf{.0601*}                 & 8.68\%                                      \\ \bottomrule
		\end{tabular}}
	\end{table*}
	
	Table \ref{tab:OverallComp} summarizes the overall performance comparison across three cross-domain scenarios, where the last column ``Imp.'' denotes the relative improvement of our method over the best-performing baseline. Our proposed KGBridge achieves consistent performance gains across all scenarios and most evaluation metrics, demonstrating the effectiveness of integrating relation-level KG semantics into sequential recommendation under non-overlapping settings. While statistical significance is observed in most cases, minor variations occur in a few settings due to differences in data sparsity and cross-domain distribution shifts. We summarize the key observations as follows:

	\paragraph{Effectiveness of KG-enhanced sequential modeling} Traditional KG-based methods, such as \textbf{KGAT} and \textbf{KMGCDR}, focus solely on structural semantics and ignore the temporal evolution of user behaviors, leading to significantly inferior performance compared to KGBridge. This demonstrates that effectively combining sequential dependencies with KG semantics can substantially enhance cross-domain transfer performance. By integrating relation-guided prompts into sequential encoders, KGBridge effectively bridges the gap between static KG relations and dynamic user preferences, yielding significant improvements over purely KG-based approaches.
	
	\paragraph{Advantages over single-domain sequential models} Compared with pure sequential baselines such as \textbf{SRGNN} and \textbf{GRU4Rec}, KGBridge achieves consistent improvements of 30.89\% and 15.02\% on average, highlighting the complementary role of KG relations in enriching item semantics and alleviating data sparsity. In contrast, oversimplified fusion of KG entities, as in \textbf{GRU4RecKG}, leads to limited or even negative gains especially on the Facebook dataset, demonstrating that directly injecting long-tailed entity embeddings may introduce noise. Among KG-aware sequential models, \textbf{KSR} incorporates both entities and relations through a key–value memory network but its performance remains sensitive to KG sparsity, resulting in unstable effect (see Section~\ref{sec:sparsity}). KGBridge consistently outperforms KSR by 50.89\% on average, demonstrating stronger robustness under sparse and incomplete KG settings. Compared with strong sequential baselines without KG enhancement, including \textbf{SASRec}, KGBridge still yields a consistent performance gain (10.22\% on average), indicating that the improvements are not solely attributable to the sequence encoder but stem from effective relation-guided knowledge integration. 
	
	\paragraph{Cross-domain adaptability}
	\textbf{KMGCDR} and \textbf{MCRPL} address cross-domain adaptation from complementary perspectives: KMGCDR exploits KG semantics but relies on user overlap, whereas MCRPL enables non-overlapping transfer via prompt-based pretraining without KG guidance. By integrating relation-level relational priors from KGs into a prompt learning framework, our  KGBridge unifies these advantages and achieves more robust cross-domain adaptation. Consequently, it yields an average improvement of 9.22\% over MCRPL and notably surpasses KMGCDR across all metrics, demonstrating that structured relational guidance effectively enhances semantic alignment and transfer stability. Compared to the generative baseline \textbf{LLM4CDSR}, KGBridge achieves a substantial performance gain across all tasks. This gap suggests the limitations of generative semantic alignment in identity-isolated settings. Specifically, LLM4CDSR leverages large language models to generate unified user profiles from item metadata, thereby capturing cross-domain semantic priors and transferring user interests across heterogeneous domains. In addition, it employs cross-domain regularization mechanisms (e.g., contrastive alignment) to calibrate the shared representation space. However, in non-overlapping scenarios, the effectiveness of such alignment-based constraints is weakened due to the absence of explicit behavioral correspondences. As a result, the model tends to rely more heavily on LLM-generated textual profiles, which may bias the learned representations toward coarse-grained semantic summaries and reduce discriminative power for fine-grained preferences in the target domain. We further analyze this phenomenon in Section~\ref{sec:discussion}.
	
	\paragraph{Cross-platform Transfer.}
	We further examine whether the transferability of KGBridge extends to the cross-platform setting. Table~\ref{tab:cross_platform} reports the results on FB Music$\rightarrow$AM Books, while the target-only SASRec results and the within-platform AM Movie$\rightarrow$AM Books results in Table~\ref{tab:OverallComp} provide two complementary references.

	First, KGBridge consistently outperforms the target-only SASRec across all eight metrics, with relative improvements ranging from 11.66\% to 27.71\%. This indicates that incorporating knowledge from the Facebook source domain provides additional benefit beyond learning solely from the sparse Amazon Books interactions. Second, KGBridge also substantially outperforms both cross-domain baselines. Compared with the stronger baseline MCRPL, the improvements range from 41.46\% to 70.66\% across Recall and NDCG, demonstrating a clear advantage of relation-guided transfer when user-level correspondence is unavailable across platforms.
	
	Moreover, compared with the within-platform AM Movie$\rightarrow$AM Books task using the same target domain, KGBridge maintains comparable performance under cross-platform transfer. The differences remain within 6.5\% across all metrics, with Recall@20 even showing a slight increase of 0.87\%. In particular, Recall@10 and NDCG@10 decrease by only 3.92\% and 4.67\%, respectively. These results suggest that the relational knowledge captured by KGBridge remains effectively transferable when the source behavioral data originate from a different platform, providing further evidence of its robustness under strict non-overlapping conditions.
	
	\begin{table*}[t]
		\centering
		\caption{Cross-platform performance on FB Music$\rightarrow$AM Books.}
		\label{tab:cross_platform}
		\setlength{\tabcolsep}{3.5pt}
		\renewcommand{\arraystretch}{1.08}
		\resizebox{\textwidth}{!}{
			\begin{tabular}{@{}lcccccccc@{}}
				\toprule
				\multirow{2}{*}{\textbf{Method}}
				& \multicolumn{4}{c}{\textbf{Recall}}
				& \multicolumn{4}{c}{\textbf{NDCG}} \\
				\cmidrule(lr){2-5}\cmidrule(l){6-9}
				& \textbf{@3}
				& \textbf{@5}
				& \textbf{@10}
				& \textbf{@20}
				& \textbf{@3}
				& \textbf{@5}
				& \textbf{@10}
				& \textbf{@20} \\
				\midrule
				\textit{MCRPL}
				& 0.0225 $\pm$ 0.0018
				& 0.0351 $\pm$ 0.0020
				& 0.0607 $\pm$ 0.0036
				& 0.0984 $\pm$ 0.0057
				& 0.0167 $\pm$ 0.0013
				& 0.0219 $\pm$ 0.0014
				& 0.0301 $\pm$ 0.0019
				& 0.0396 $\pm$ 0.0023 \\
				
				\textit{LLM4CDSR}
				& 0.0082 $\pm$ 0.0007
				& 0.0118 $\pm$ 0.0009
				& 0.0187 $\pm$ 0.0012
				& 0.0295 $\pm$ 0.0037
				& 0.0062 $\pm$ 0.0007
				& 0.0077 $\pm$ 0.0007
				& 0.0099 $\pm$ 0.0002
				& 0.0126 $\pm$ 0.0009 \\
				
				\textit{KGBridge}
				& \textbf{0.0383 $\pm$ 0.0014}
				& \textbf{0.0569 $\pm$ 0.0018}
				& \textbf{0.0906 $\pm$ 0.0007}
				& \textbf{0.1392 $\pm$ 0.0017}
				& \textbf{0.0285 $\pm$ 0.0014}
				& \textbf{0.0361 $\pm$ 0.0014}
				& \textbf{0.0469 $\pm$ 0.0009}
				& \textbf{0.0591 $\pm$ 0.0008} \\
				\bottomrule
			\end{tabular}
		}
	\end{table*}

\paragraph{Runtime Efficiency}
To provide a more comprehensive efficiency comparison, Table~\ref{tab:runtime} reports the average per-epoch time, completed epochs, and convergence time of each method. Convergence time is measured from the start of training until early stopping. For KGBridge, Stage~I and Stage~II are measured separately, and the total convergence time is computed from the unrounded stage-wise measurements. Traditional sequential models, including \textbf{SASRec}, \textbf{SRGNN}, and \textbf{GRU4Rec}, generally converge faster due to their lightweight architectures and target-only training. In comparison, KG-enhanced methods incur additional computational costs in modeling structured semantics.

Among the cross-domain methods, KGBridge converges in 54.55, 90.41, and 5.35 minutes on the three tasks, respectively. The corresponding Stage~I/Stage~II times are 49.94/4.62, 83.78/6.63, and 4.30/1.05 minutes, showing that Stage~I accounts for 80.42\%--92.67\% of the total training time. KGBridge requires considerably less convergence time than \textbf{KMGCDR} across all three tasks. Compared with \textbf{LLM4CDSR}, it shows comparable efficiency on FB Movie$\rightarrow$Books and converges faster on the other two tasks. Although \textbf{MCRPL} terminates sooner, KGBridge consistently achieves higher recommendation performance, as reported in Table~\ref{tab:OverallComp}.

Overall, KGBridge maintains competitive training efficiency while incorporating structured relational semantics for cross-domain transfer. The stage-wise results further show that its computational cost is concentrated in joint cross-domain pretraining, followed by lightweight target-domain adaptation.

\begin{table*}[t]
	\caption{Training efficiency comparison.}
	\label{tab:runtime}
	\centering
	\setlength{\tabcolsep}{3.5pt}
	\renewcommand{\arraystretch}{1.08}
	\resizebox{\textwidth}{!}{
		\begin{tabular}{@{}llcccccccccc@{}}
			\toprule
			\multirow{2}{*}{\textbf{Dataset}}
			& \multirow{2}{*}{\textbf{Measure}}
			& \multicolumn{6}{c}{\textbf{Single-domain Methods}}
			& \multicolumn{4}{c}{\textbf{Cross-domain Methods}} \\
			\cmidrule(lr){3-8}\cmidrule(l){9-12}
			&
			& \textit{SASRec}
			& \textit{SRGNN}
			& \textit{GRU4Rec}
			& \textit{GRU4RecKG}
			& \textit{KGAT}
			& \textit{KSR}
			& \textit{KMGCDR}
			& \textit{LLM4CDSR}
			& \textit{MCRPL}
			& \textit{KGBridge} \\
			\midrule
			\multirow{3}{*}{\shortstack[l]{FB-Movie\\$\rightarrow$ Books}}
			& Per-epoch time (s)
			& 15.34 & 32.00 & 6.66 & 2.96 & 27.17 & 14.32
			& 490.13 & 251.31 & 17.37 & 22.89 \\
			& Completed epochs
			& 98 & 42 & 100 & 100 & 53 & 68
			& 40 & 13 & $46+54$ & $93+50$ \\
			& Convergence time (min)
			& 25.05 & 22.40 & 11.10 & 4.93 & 24.00 & 16.23
			& 326.75 & 54.45 & 28.95 & 54.55 \\
			\midrule
			\multirow{3}{*}{\shortstack[l]{FB-Music\\$\rightarrow$ Books}}
			& Per-epoch time (s)
			& 15.34 & 32.00 & 6.66 & 7.18 & 54.26 & 15.60
			& 472.39 & 798.11 & 25.61 & 32.29 \\
			& Completed epochs
			& 98 & 42 & 100 & 100 & 58 & 89
			& 76 & 34 & $55+67$ & $97+71$ \\
			& Convergence time (min)
			& 25.05 & 22.40 & 11.10 & 11.97 & 52.45 & 23.13
			& 598.37 & 452.26 & 52.07 & 90.41 \\
			\midrule
			\multirow{3}{*}{\shortstack[l]{AM-Movie\\$\rightarrow$ Books}}
			& Per-epoch time (s)
			& 7.75 & 10.07 & 1.69 & 1.80 & 16.49 & 3.25
			& 65.43 & 60.16 & 3.51 & 3.65 \\
			& Completed epochs
			& 56 & 30 & 83 & 59 & 51 & 52
			& 49 & 25 & $52+17$ & $65+23$ \\
			& Convergence time (min)
			& 7.23 & 5.03 & 2.33 & 1.77 & 14.02 & 2.82
			& 53.43 & 25.07 & 4.04 & 5.35 \\
			\bottomrule
		\end{tabular}
	}
	\vspace{2pt}
	\begin{minipage}{\textwidth}
		\footnotesize
		\textit{Note:} For two-stage methods, $a+b$ denotes the completed epochs in Stage~I and Stage~II, respectively.
	\end{minipage}
\end{table*}

	\subsection{Analysis of Model Design (Q2)}\label{sec:model_analysis}
	To examine the design of KGBridge, we conduct two complementary analyses. First, we evaluate the contribution of each key component through component-wise ablations. Second, we isolate the effect of the KG representation unit through a controlled comparison between relation-level and entity-level inputs, while keeping the prompt architecture, training procedure, regularization, and downstream parameter budget unchanged.

	\subsubsection{Component-wise ablation.} We selectively modify or disable modules in the proposed framework. Specifically, we evaluate the following variants: 
	\textit{-KGInit} replaces relation-guided prompt initialization with Xavier normal initialization;
	\textit{-Shared} replaces domain-shared prompts with Xavier-initialized ones\footnote{Using the replace operation instead of removal is to maintain parameter dimension};
	\textit{-Spec} replaces domain-specific prompts with Xavier-initialized ones;
	\textit{-Disen} removes the disentanglement regularization term; and 
	\textit{-Freeze} updates all prompts during fine-tuning instead of freezing shared ones. 
	
		\begin{figure}[]
		\centering
		\includegraphics[width=\linewidth]{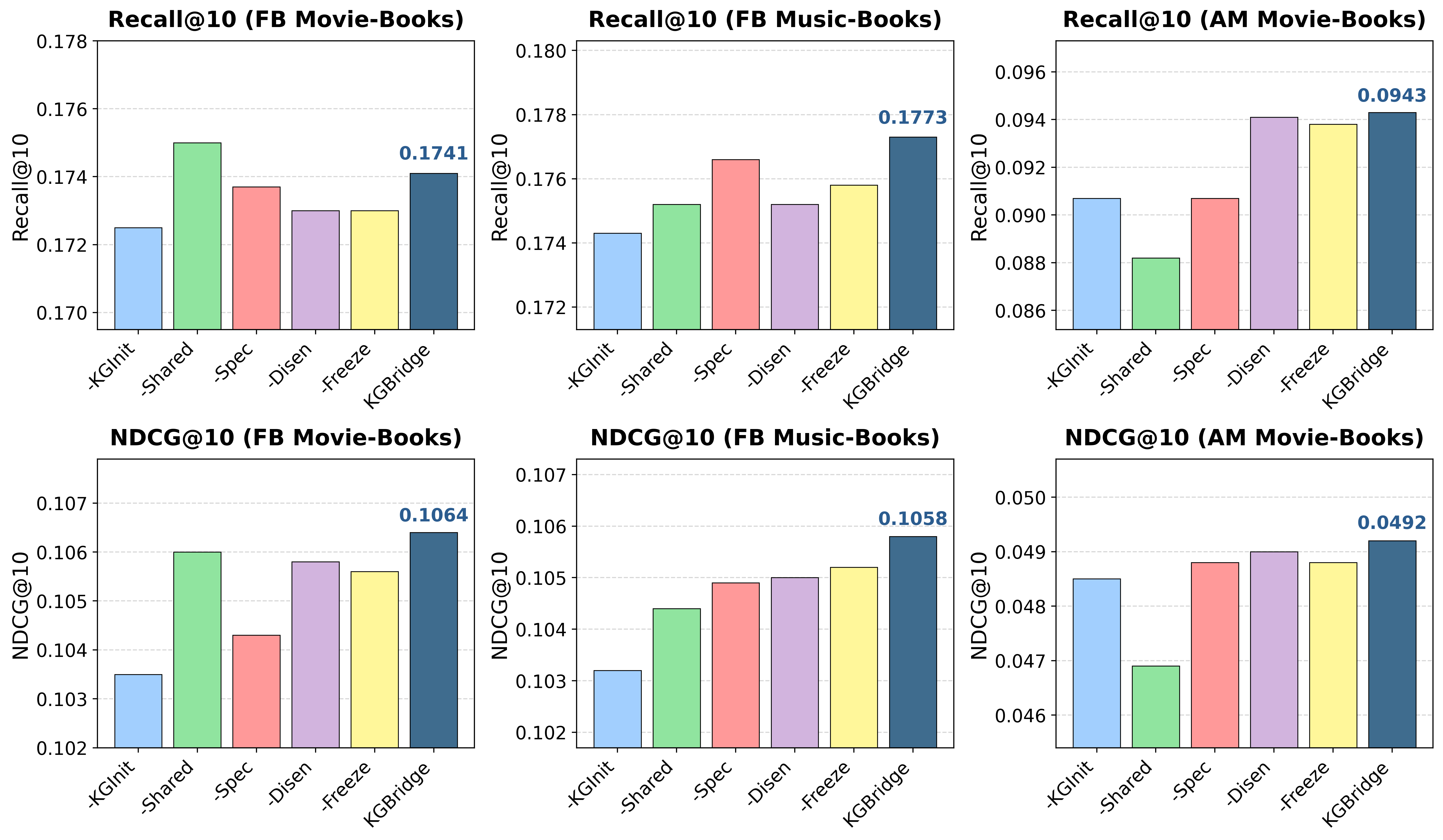}
		\caption{Component-wise ablation of KGBridge across three transfer scenarios in terms of Recall@10 and NDCG@10.}
		\label{fig:ablation_grid}
	\end{figure}

	The comprehensive comparison across three scenarios is illustrated in Fig.~\ref{fig:ablation_grid}. We have three key observations:
	(1) Removing relation-guided initialization (\textit{-KGInit}) or replacing shared prompts (\textit{-Shared}) leads to the largest performance drops (2.23\% and 2.37\% on average, respectively), highlighting the importance of structured relational semantics and domain-invariant representations for effective cross-domain transfer. Notably, on the sparse AM~Movie-Books dataset, replacing the KG-guided shared prompts results in a sharp Recall@10 drop of 6.47\%, which further underscores the necessity of anchoring domain-invariant knowledge in relational priors.
	(2) The degradation observed for \textit{-Spec} (1.38\% on average) demonstrates that domain-specific prompts provide complementary adaptation by capturing fine-grained domain nuances beyond shared priors, ensuring that the model can effectively bridge the domain-specific semantic gap.
	(3) The smaller gaps associated with \textit{-Disen} and \textit{-Freeze}, which operate only during target-domain adaptation, suggest that these mechanisms provide complementary refinement on top of the transferable relational priors established during joint pre-training. We further analyze the learned prompt representations and adaptation behaviors in Section~\ref{sec:prompt_analysis}.
	
	\subsubsection{Relation-level versus entity-level KG inputs.}
	To isolate the effect of the KG representation unit, we replace relation embeddings with entity embeddings learned under the same TransE training setup, while retaining the identical prompt construction and optimization pipeline. Table~\ref{tab:relation_entity} reports the mean and standard deviation across runs; the KG input count refers to the number of embeddings before fixed-length prompt generation.
	
	\begin{table}[t]
		\centering
		\caption{Controlled comparison of relation-level and entity-level KG inputs under identical KGBridge architecture and training settings.}
		\label{tab:relation_entity}
		\setlength{\tabcolsep}{4pt}
		\begin{tabular}{llrrr}
			\toprule
			\textbf{Task} & \textbf{KG Unit} & \textbf{\# KG Inputs}
			& \textbf{Recall@10} & \textbf{NDCG@10} \\
			\midrule
			\multirow{2}{*}{FB Movie$\rightarrow$Books}
			& Entity   & 41,768  & 0.1737 $\pm$ 0.0039 & 0.1046 $\pm$ 0.0025 \\
			& Relation & 32      & \textbf{0.1740 $\pm$ 0.0008} & \textbf{0.1049 $\pm$ 0.0013} \\
			\midrule
			\multirow{2}{*}{FB Music$\rightarrow$Books}
			& Entity   & 148,017 & 0.1753 $\pm$ 0.0018 & 0.1058 $\pm$ 0.0011 \\
			& Relation & 32      & \textbf{0.1770 $\pm$ 0.0022} & \textbf{0.1063 $\pm$ 0.0013} \\
			\midrule
			\multirow{2}{*}{AM Movie$\rightarrow$Books}
			& Entity   & 87,234  & 0.0909 $\pm$ 0.0019 & 0.0481 $\pm$ 0.0010 \\
			& Relation & 23      & \textbf{0.0921 $\pm$ 0.0016} & \textbf{0.0486 $\pm$ 0.0006} \\
			\bottomrule
		\end{tabular}
	\end{table}

	Relation-level inputs achieve consistently higher mean Recall@10 and NDCG@10 across all three scenarios, while reducing the raw KG input from 41,768--148,017 entity embeddings to only 23--32 relation embeddings, yielding a reduction of more than three orders of magnitude in the KG input size before prompt generation. Since all downstream components are controlled, this comparison isolates the effect of the KG representation unit and shows that relation-level abstraction provides a substantially more compact semantic basis while maintaining higher mean recommendation performance. This supports our motivation that relations capture recurring cross-domain semantics without relying on the large and long-tailed entity space.
	
	Together, the two analyses validate both the individual components of KGBridge and the use of relation-level semantics as the basis for knowledge-guided prompt learning.

	\subsection{Cross-domain Consistency of Shared Relations (Q3)}
	As introduced in Section~\ref{sec:prompt_agg}, KGBridge aggregates domain-shared relation embeddings to construct shared prompts for cross-domain knowledge transfer. In the DBpedia-based KGs, shared relations are identified through exact relation-identifier intersection, where identical identifiers refer to the same DBpedia property across domains. However, the same relation may exhibit different structural usage patterns in different domains. We therefore examine the cross-domain consistency of these shared relations through a co-occurrence analysis.
	
	Specifically, for each shared relation $r$, we construct a relation co-occurrence profile in each domain. For every other relation type $q$, the corresponding profile entry records the number of distinct head entities on which $r$ and $q$ both occur. The entry corresponding to $r$ itself is excluded to avoid trivial matching, and the profiles are constructed over the union of the source- and target-domain relation vocabularies. We then compare the source- and target-domain profiles using cosine similarity. Shared relations with an all-zero profile in either domain are excluded because their cosine similarity is undefined.
	
	Table~\ref{tab:relation_consistency} reports three complementary statistics. \textit{Same-ID} measures the average similarity between source- and target-domain profiles with the same relation identifier. \textit{Different-ID} averages the similarities over all cross-domain profile pairs with different identifiers among the remaining shared relations. \textit{Top-3 Match} reports the proportion of analyzed shared relations whose Same-ID counterpart ranks among the three most similar target-domain profiles within the same relation set.

	\begin{table}[t]
		\centering
		\caption{Cross-domain consistency of relation co-occurrence profiles.}
		\label{tab:relation_consistency}
		\setlength{\tabcolsep}{4pt}
		\begin{tabular}{lrrrr}
			\toprule
			\textbf{Task} & \textbf{Analyzed/Shared}
			& \textbf{Same-ID} & \textbf{Different-ID}
			& \textbf{Top-3 Match} \\
			\midrule
			FB Movie$\rightarrow$Books & 21/25 & 0.734 & 0.269 & 76.2\% \\
			FB Music$\rightarrow$Books & 20/24 & 0.366 & 0.173 & 45.0\% \\
			AM Movie$\rightarrow$Books & 17/17 & 0.742 & 0.503 & 64.7\% \\
			\bottomrule
		\end{tabular}
	\end{table}
	
	As shown in Table~\ref{tab:relation_consistency}, Same-ID profile pairs exhibit higher cross-domain co-occurrence similarity than Different-ID pairs in all three tasks, indicating that relations with the same identifier tend to exhibit more similar structural usage patterns across domains. This pattern is particularly clear for the two Movie$\rightarrow$Books tasks, where Same-ID similarity reaches 0.734 and 0.742, and 76.2\% and 64.7\% of the analyzed relations have their Same-ID counterparts ranked within the top three, respectively. FB Music$\rightarrow$Books shows weaker consistency, with a Same-ID similarity of 0.366 and a Top-3 Match rate of 45.0\%, suggesting greater variation in relation usage across this domain pair. Nevertheless, its Same-ID similarity remains higher than the Different-ID similarity of 0.173.
	
	Overall, these results show that relations identified by exact relation-identifier intersection generally exhibit more similar cross-domain usage patterns than relations with different identifiers. The variation across tasks also indicates that such consistency is not uniform, supporting the joint modeling of shared and domain-specific relational semantics in KGBridge.
	
	\subsection{Impact of Prompt Aggregation (Q4)} \label{sec:agg_study}
	To investigate how different aggregation strategies affect the quality of relation-guided prompt construction, we compare four variants: \textit{mean}, \textit{attention}, \textit{transformer}, and the proposed \textit{mean\_noise}. The corresponding formulations are introduced in Section~\ref{sec:prompt_gen_impl}. Results are reported in Table~\ref{tab:agg_ablation}.

\begin{table}[]
	\centering
	\caption{Performance comparison of different prompt aggregation strategies.}
	\label{tab:agg_ablation}
	\resizebox{\linewidth}{!}{
		\begin{tabular}{l|cccc|cccc|cccc}
			\toprule
			& \multicolumn{4}{c|}{FB Movie-Books} & \multicolumn{4}{c|}{FB Music-Books} & \multicolumn{4}{c}{AM Movie-Books} \\
			Strategy & R@5 & R@10 & N@5 & N@10 & R@5 & R@10 & N@5 & N@10 & R@5 & R@10 & N@5 & N@10 \\
			\midrule
			mean        & 0.1256 & 0.1732 & 0.0896 & 0.1049 & \textbf{0.1280} & 0.1758 & \textbf{0.0907} & 0.1061 & 0.0586 & 0.0924 & 0.0379 & 0.0488 \\
			attention   & 0.1251 & 0.1736 & 0.0896 & 0.1052 & 0.1235 & 0.1711 & 0.0877 & 0.1031 & 0.0572 & 0.0900 & 0.0376 & 0.0481 \\
			transformer & 0.1249 & 0.1732 & 0.0894 & 0.1051 & 0.1274 & \textbf{0.1782} & 0.0904 & \textbf{0.1067} & 0.0578 & 0.0907 & 0.0378 & 0.0484 \\
			\rowcolor{gray!20} mean\_noise  & \textbf{0.1275} & \textbf{0.1741} & \textbf{0.0914} & \textbf{0.1064} & 0.1273 & 0.1773 & 0.0898 & 0.1058 & \textbf{0.0608} & \textbf{0.0943} & \textbf{0.0383} & \textbf{0.0492} \\
			\bottomrule
	\end{tabular}}
\end{table}

We derive the following observations from the results.

\textbf{(1) Effectiveness of Stochastic Aggregation.} 
The proposed \textit{mean\_noise} consistently achieves the best overall performance on FB~Movie-Books and AM~Movie-Books across all evaluation metrics. Compared with deterministic aggregation, stochastic perturbation encourages the prompt generator to focus on invariant relation-level semantics rather than memorizing specific relation co-occurrences. This implicit regularization effect improves robustness and generalization under sparse and structurally imbalanced cross-domain supervision, validating the theoretical motivation discussed in Section~\ref{sec:prompt_gen_impl}.

\textbf{(2) Trade-off Between Simplicity and Expressiveness.}
On FB~Music-Books, different aggregation strategies exhibit distinct advantages under different ranking depths. Specifically, \textit{mean} pooling achieves competitive performance on top-ranked metrics (R@5 and N@5), while \textit{transformer}-based aggregation performs better at larger cutoffs (R@10 and N@10). This phenomenon suggests a trade-off between semantic simplicity and structural expressiveness. Simpler pooling mechanisms effectively preserve dominant semantic signals for highly confident recommendations, whereas transformer-based aggregation better captures higher-order relational dependencies that benefit deeper retrieval.

\textbf{(3) Limitations of Highly Expressive Aggregators.}
Although attention and transformer-based strategies provide stronger modeling capacity, they do not consistently outperform lightweight aggregation methods across datasets. This indicates that increased aggregation complexity does not necessarily improve prompt quality under sparse relation distributions. In practice, over-parameterized aggregators may introduce optimization instability and amplify noisy semantic correlations, leading to inferior generalization compared with simpler strategies.

Overall, the experimental results demonstrate that relation-guided prompt construction benefits more from controlled stochastic regularization than from excessive aggregation complexity. By balancing semantic stability and representation diversity, \textit{mean\_noise} provides a robust and efficient mechanism for cross-domain prompt generation under sparse KG supervision.

	\subsection{Effect of KG Sparsity (Q5)} \label{sec:sparsity}
\begin{figure}[]
		\centering
		\includegraphics[width=0.9\linewidth]{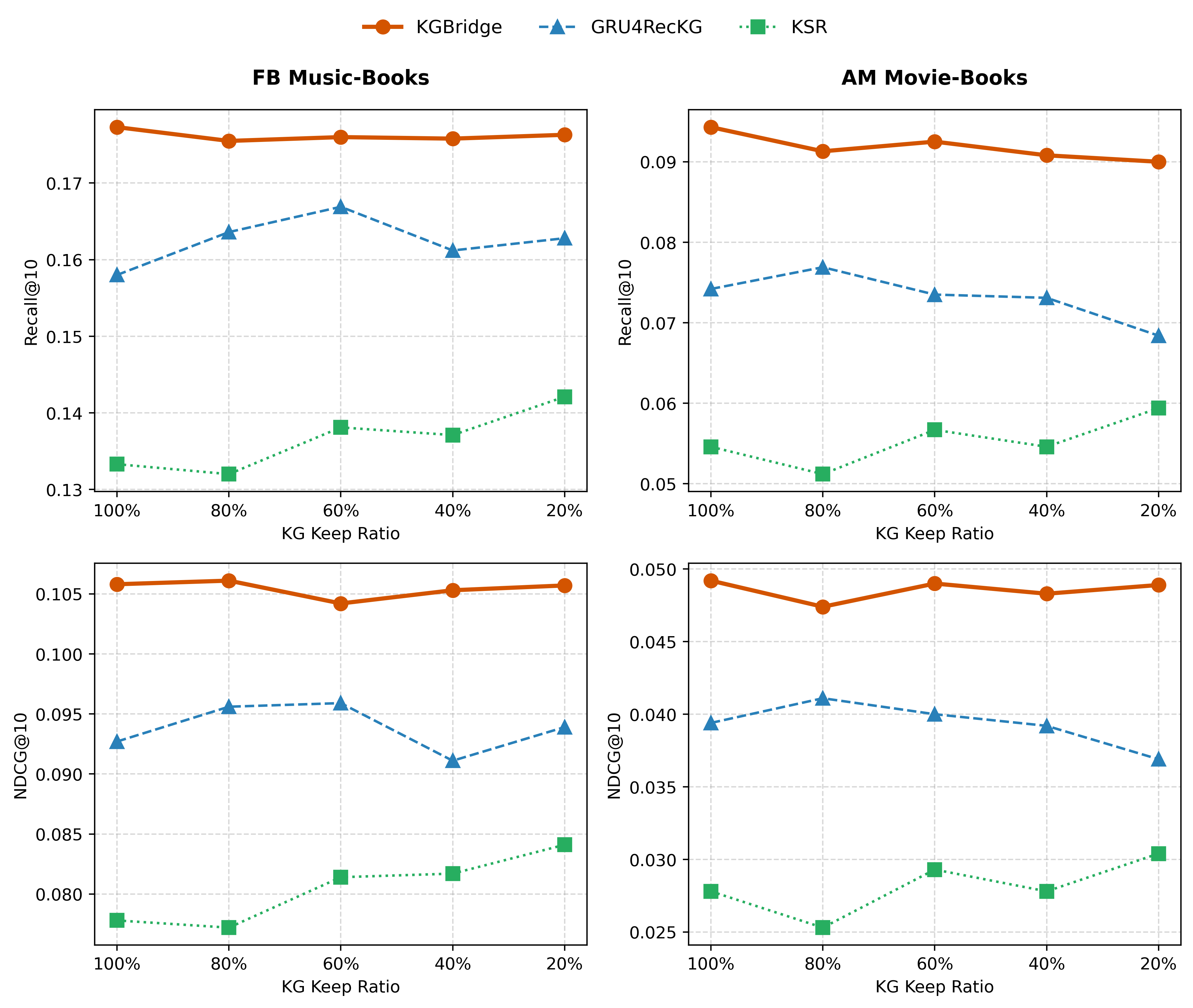}
		\caption{Effect of KG sparsity on KG-based methods.}
		\label{fig:sparsity}
	\end{figure}

	To evaluate robustness under incomplete knowledge graphs, we simulate KG sparsity by randomly removing 20\%, 40\%, 60\%, and 80\% of triples from the original KG. Although real-world KGs often exhibit long-tail distributions and structural imbalance, such properties are difficult to control directly in experiments. Therefore, varying KG sparsity provides a practical alternative for assessing robustness under incomplete and uneven KG conditions. 
	We compare KGBridge with two representative KG-based baselines: \textbf{GRU4RecKG}, which leverages only entity embeddings for sequential modeling, and \textbf{KSR}, which models entities and relations via a key–value memory network. Since all cross-domain tasks exhibit consistent trends, we report representative results on FB~Music-Book and AM~Movie-Book, as illustrated in Fig.~\ref{fig:sparsity}.

		We summarize the key observations from this experiment as follows:
	
	(1) \textbf{Superiority under Sparse Conditions.} KGBridge consistently outperforms both baselines across all sparsity levels, demonstrating strong robustness to KG incompleteness. For instance, on the AM~Movie-Books task, KGBridge improves Recall@10 over GRU4RecKG by 27.09\%, 18.73\%, 25.85\%, 24.21\%, and 31.58\% under increasing sparsity levels. Notably, even with only 20\% of triples retained, KGBridge maintains competitive performance, indicating that relation-guided prompts capture condensed and semantically stable patterns that are less sensitive to missing structural information.
	
	(2) \textbf{Robustness of Relational Priors.} As sparsity increases, GRU4RecKG and KSR exhibit pronounced performance fluctuations, whereas KGBridge shows significantly smoother trends. This contrast highlights that relation-level semantic priors provide a more reliable foundation for cross-domain transfer than entity-centric methods, which are highly sensitive to KG information density and local structural changes.
	
	(3) \textbf{Noise Sensitivity in Baselines.} Notably, KSR exhibits occasional performance gains as KG sparsity increases. This counterintuitive effect likely stems from the incidental removal of redundant or noisy triples, which can temporarily improve representation quality. However, this trend also reveals the high sensitivity of KSR to KG changes, with performance fluctuating substantially across settings. In contrast, KGBridge maintains consistently stable accuracy, demonstrating its ability to leverage core relational semantics effectively even under varying KG conditions.
	
	\subsection{Convergence Analysis (Q6)} \label{sec:convergence}
\begin{figure}[]
		\centering
		\includegraphics[width=1\linewidth]{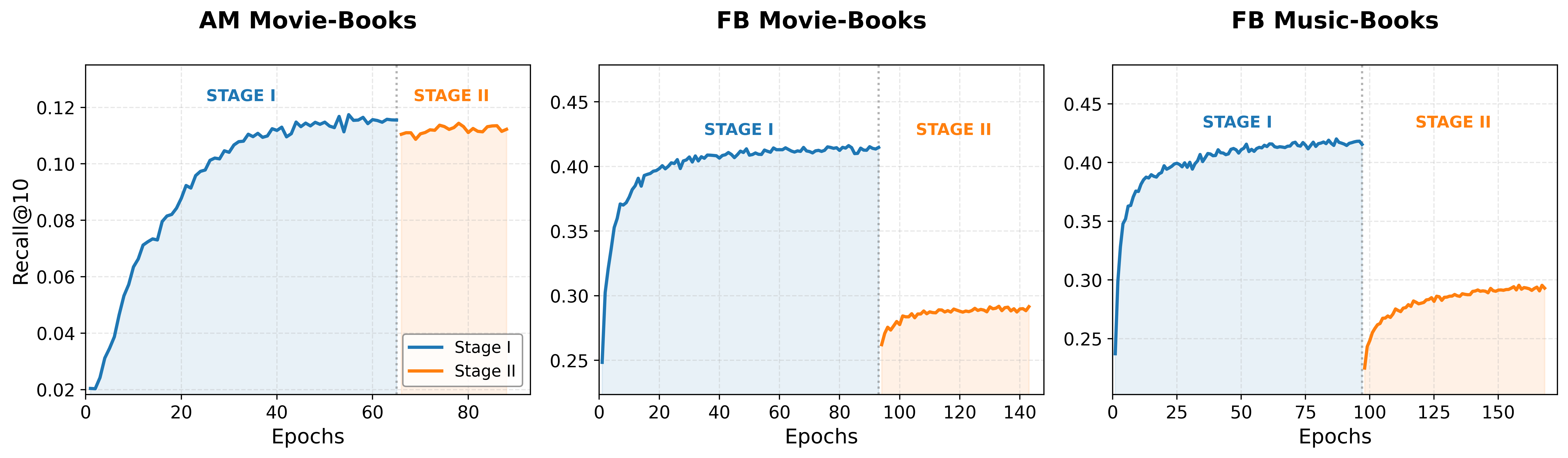}
		\caption{Convergence analysis of KGBridge across different scenarios. Stage I represents joint pre-training on all domains, and Stage II represents target-only adaptation.}
		\label{fig:convergence}
	\end{figure}

	To evaluate the optimization stability and adaptation effectiveness of KGBridge, we visualize the convergence behavior under three representative cross-domain scenarios. As shown in Fig.~\ref{fig:convergence}, KGBridge exhibits consistent convergence patterns across both pretraining and target adaptation stages, demonstrating stable optimization and effective knowledge transfer.

	We summarize four main observations as follows. (1) \textbf{Rapid Transferable Representation Learning.}  During \textit{Stage I (Joint Pre-training)}, Recall@10 increases rapidly within the first 20 epochs across all scenarios. This indicates that relation-guided prompts provide strong semantic priors that facilitate efficient cross-domain representation learning and enable the model to quickly establish transferable semantic structures. (2) \textbf{Effectiveness of Two-Stage Adaptation.}  When transitioning from \textit{Stage I} to \textit{Stage II (Target Adaptation)}, a temporary performance fluctuation is observed due to the shift from joint-domain optimization to target-specific adaptation. Nevertheless, performance consistently improves during Stage II, demonstrating that freezing shared prompts while adapting domain-specific prompts effectively reduces residual domain divergence and enhances target-domain alignment. (3) \textbf{Robustness Across Different Scenarios.} Despite substantial differences in interaction sparsity and KG density between Amazon and Facebook datasets, KGBridge maintains highly consistent convergence behaviors across all settings. This suggests that the proposed framework generalizes well under diverse cross-domain environments. (4) \textbf{Stable Optimization Behavior}: 
	Both training stages converge smoothly to stable performance plateaus before early stopping, indicating that the disentangled prompt design and contrastive regularization effectively stabilize optimization and mitigate overfitting during cross-domain transfer.
	
	\subsection{Prompt Disentanglement Analysis (Q7)}
	\label{sec:prompt_analysis}
	To further examine how the Shared and Specific prompts are organized during target-domain adaptation, we analyze the learned prompts at both the representation and parameter levels. The Shared prompts are constructed from relations occurring in both domains, while the Specific prompts are constructed from the remaining domain-dependent relations. The two prompt banks therefore originate from separated relational sources and play different roles during Stage~II. We examine whether they exhibit distinct representation structures, whether the same-index correspondence encouraged by $\mathcal{L}_{\text{disen}}$ emerges after adaptation, and whether their parameter updates follow the intended shared-stable and specific-adaptive behaviors.
	
\begin{figure}[]
		\centering
		\includegraphics[width=\linewidth]{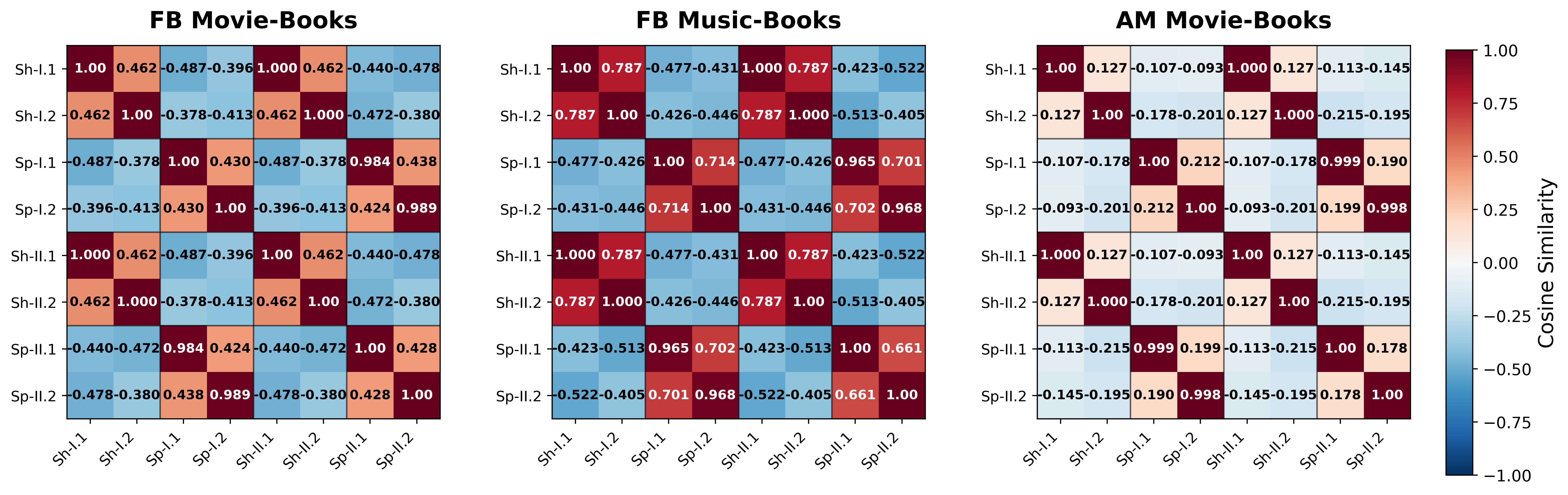}
		\caption{Pairwise cosine similarity between prompt vectors across Stage~I and Stage~II. ``Sh'' and ``Sp'' denote Shared and Specific prompts, respectively. Within each stage-specific Shared--Specific block, diagonal cells correspond to same-index pairs and off-diagonal cells to cross-index pairs.}
		\label{fig:prompt_cosine}
	\end{figure}

	\paragraph{Representation-level Analysis via Cosine Similarity}
	We extract the Shared and Specific prompts from the checkpoints obtained after Stage~I and Stage~II and compute pairwise cosine similarities among all prompt vectors. Each prompt bank contains $L=2$ vectors with dimension $d=100$. Consequently, each transfer scenario contains eight prompt vectors across the two stages, producing the $8\times8$ similarity matrices shown in Fig.~\ref{fig:prompt_cosine}.
	
	To examine the correspondence structure, we focus on the within-stage Shared--Specific blocks. For each stage, $(\mathrm{Sh.1},\mathrm{Sp.1})$ and $(\mathrm{Sh.2},\mathrm{Sp.2})$ form same-index pairs, while $(\mathrm{Sh.1},\mathrm{Sp.2})$ and $(\mathrm{Sh.2},\mathrm{Sp.1})$ form cross-index pairs. We further compare their mean cosine similarities by subtracting the mean cross-index similarity from the mean same-index similarity. A positive value therefore indicates stronger correspondence between same-index prompt slots.
	
	Figure~\ref{fig:prompt_cosine} reveals three consistent patterns.
	(1) \textbf{Preservation of Shared prompts.} The Shared representations remain identical between Stage~I and Stage~II across all scenarios, confirming that the freezing operation preserves the learned shared representations during target-domain adaptation.
	(2) \textbf{Representational distinction between prompt types.} The Shared--Specific cosine similarities remain negative across both stages, indicating clear representational separation between the two prompt banks.
	(3) \textbf{Emergence of same-index correspondence.}
	For the three scenarios, the similarity difference between same-index and cross-index pairs changes from $-0.063$, $-0.033$, and $-0.019$ after Stage~I to $0.065$, $0.104$, and $0.026$ after Stage~II, respectively. The consistent change from negative to positive values indicates that same-index Shared--Specific pairs develop stronger relative correspondence than cross-index pairs after target-domain adaptation, consistent with the structure encouraged by $\mathcal{L}_{\text{disen}}$.
	
\begin{figure}[]
		\centering
		\includegraphics[width=\linewidth]{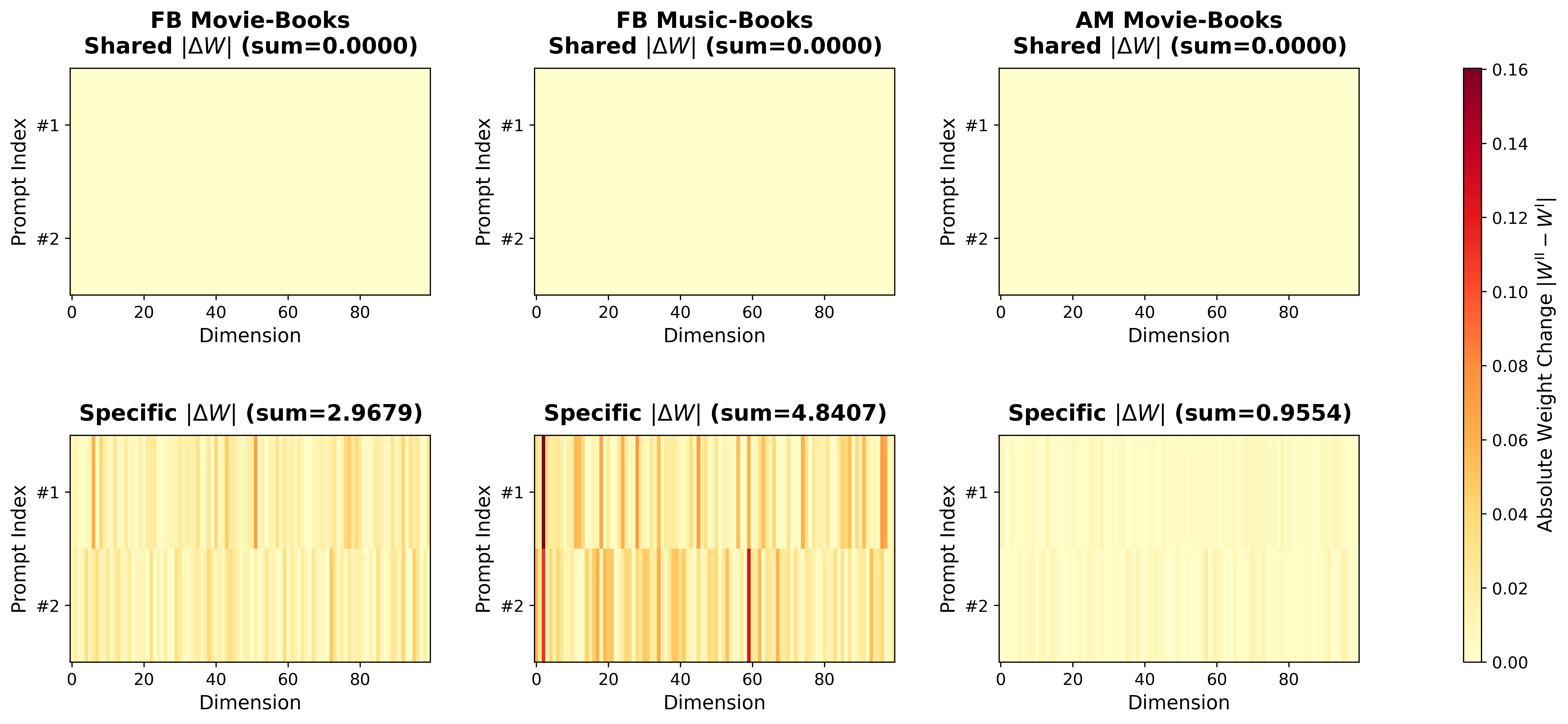}
		\caption{Dimension-wise absolute weight change $|W^{\mathrm{II}}-W^{\mathrm{I}}|$ from Stage~I to Stage~II. The upper row shows the frozen Shared prompts, while the lower row shows the target-adaptive Specific prompts.}
		\label{fig:prompt_weight}
	\end{figure}

	\paragraph{Parameter-level Analysis via Weight Dynamics}
	We further examine the adaptation behaviors of the two prompt banks by calculating the element-wise absolute difference $|W^{\mathrm{II}}-W^{\mathrm{I}}|$ between their Stage~I and Stage~II parameters. Figure~\ref{fig:prompt_weight} visualizes these differences, with the Shared prompts shown in the upper row and the Specific prompts in the lower row.
	
	The two prompt banks exhibit clearly different update behaviors.
	(1) \textbf{Stable Shared prompts.} The Shared prompts remain unchanged across all dimensions and scenarios, with a total absolute difference of $0.0000$, confirming that the learned shared relational information is preserved throughout target-domain adaptation.
	(2) \textbf{Adaptive Specific prompts.} In contrast, the Specific prompts are updated under target-domain supervision. Their total absolute changes are $2.9679$, $4.8407$, and $0.9554$, respectively, following the same task order as in Fig.~\ref{fig:prompt_weight}. The different update magnitudes show that the extent of adaptation varies across transfer scenarios, while the non-uniform changes across dimensions indicate that Stage~II adjusts different components of the Specific prompts to different degrees.
	
	Overall, the representation- and parameter-level analyses show that the Shared and Specific prompts maintain distinct representation structures while exhibiting structured same-index correspondence after target-domain adaptation. Meanwhile, the Shared prompts remain stable across stages, whereas the Specific prompts adapt to the target domain. These results support the complementary roles of shared and domain-specific relational information in the prompt disentanglement design and provide representation-level evidence consistent with the correspondence refinement encouraged by $\mathcal{L}_{\text{disen}}$.

	\subsection{Hyperparameter Sensitivity (Q8)}\label{sec:hyperparam}
	We further analyze the sensitivity of KGBridge to the balancing coefficient
	$\lambda$ in Eq.~(\ref{eq:finetune}), which controls the contribution of
	$\mathcal{L}_{\mathrm{disen}}$ during target-domain adaptation. We evaluate
	$\lambda \in \{0, 0.001, 0.002, 0.003, 0.005, 0.01, 0.1\}$ across the three
	transfer tasks, where $\lambda=0$ removes the correspondence regularization.
	The expanded range covers settings from no auxiliary constraint to
	substantially increased weighting, allowing us to examine both local
	sensitivity and the effect of stronger regularization.
	
\begin{figure}[]
		\centering
		\includegraphics[width=\linewidth]{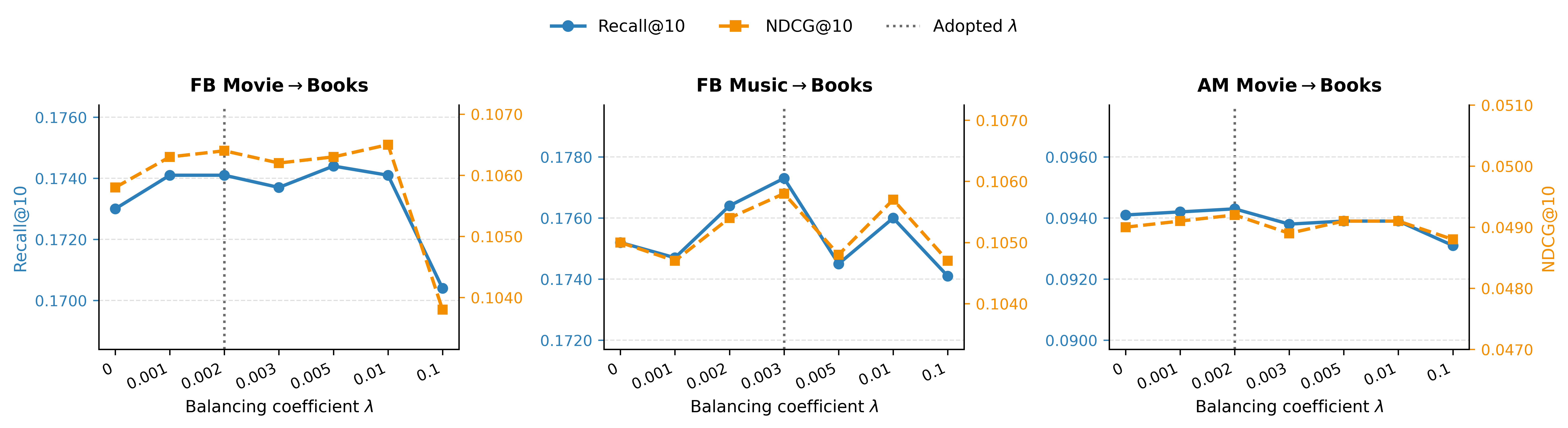}
		\caption{Comparison w.r.t. different values of the balancing coefficient $\lambda$}
		\label{fig:hyperparam}
	\end{figure}

	As shown in Fig.~\ref{fig:hyperparam}, the effect of $\lambda$ exhibits limited but task-dependent fluctuations within the range from $0.001$ to $0.01$. Across the three tasks, the maximum relative variation is $1.58\%$ for Recall@10 and $1.04\%$ for NDCG@10. Compared with $\lambda=0$, the adopted settings yield improvements of $0.21\%$--$1.20\%$ in Recall@10 and $0.41\%$--$0.76\%$ in NDCG@10. When $\lambda$ is increased to $0.1$, both metrics decrease across all
	three tasks, with the largest degradation observed on FB Movie--Books. These results indicate that correspondence regularization provides consistent benefits when assigned a moderate weight, while excessive weighting can interfere with the primary recommendation objective.
	
	\subsection{Qualitative Analysis and Discussion}\label{sec:discussion}
\begin{figure}[]
		\centering
		\includegraphics[width=0.7\linewidth]{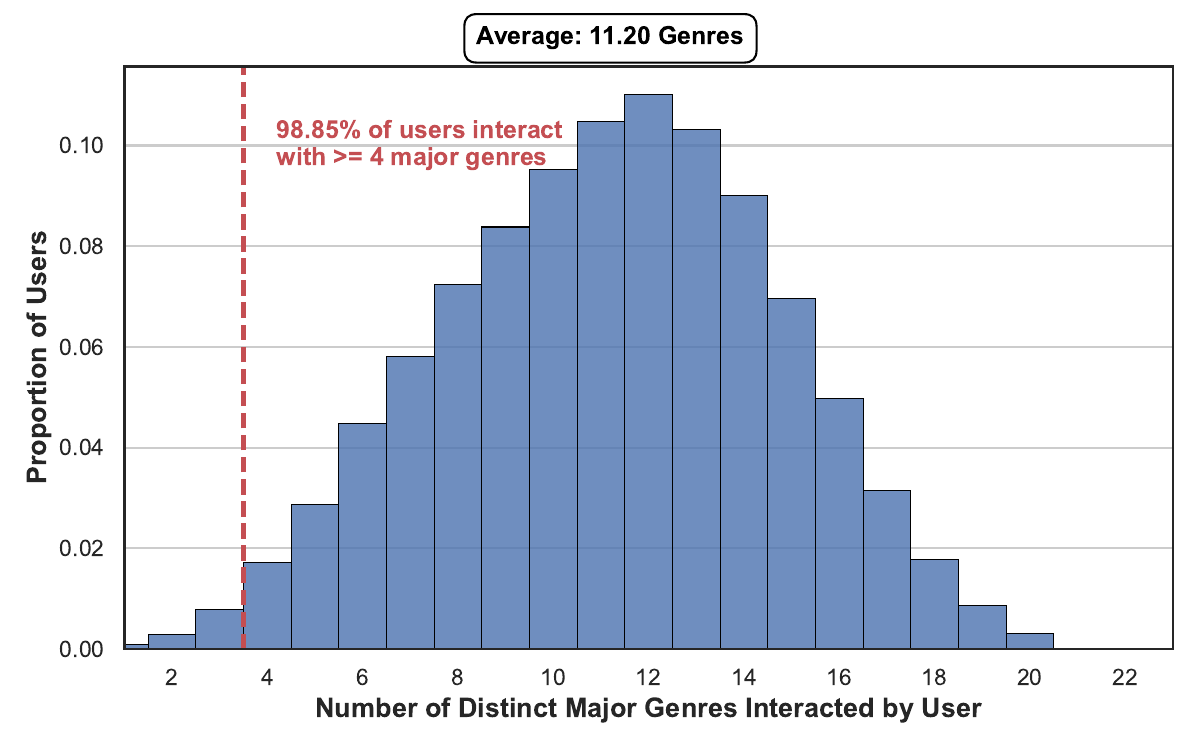}
		\caption{Empirical distribution of user semantic diversity in the FB-Movie dataset. The high prevalence of multi-genre interactions (98.85\% of users cover $\ge 4$ distinct major genres) highlights the importance of preserving fine-grained semantics under non-overlapping recommendation settings.}
		\label{fig:case_study}
	\end{figure}

	The overall empirical results in Table~\ref{tab:OverallComp}  demonstrate that KGBridge consistently outperforms LLM4CDSR under non-overlapping settings. To further examine the differences between the two semantic transfer mechanisms, we conduct a qualitative analysis using User~61353, as detailed in  Table~\ref{tab:case_user}. This user was selected because the interaction history spans at least four major genres, consistent with the multi-genre criterion used in Fig.~\ref{fig:case_study}. Since 98.85\% of users in FB-Movie satisfy this criterion, the case reflects a prevalent multi-genre interaction pattern and does not constitute an exceptional profile.


\begin{table}[]
	\centering
	\caption{Case study comparing semantic representation mechanisms and recommendation behaviors for User ID: 61353 on the FB-Movie dataset.}
	\label{tab:case_user} 
	\resizebox{\linewidth}{!}{
		\begin{tabular}{l | p{0.45\linewidth} | p{0.45\linewidth}}
			\toprule
			\textbf{User History} & \multicolumn{2}{l}{1. \textit{Pulp Fiction} (Action/Crime) \quad 2. \textit{Ray} (Drama/Biography) \quad 3. \textit{Howl's Moving Castle} (Animation/Fantasy)} ... \\ \midrule
			\textbf{Paradigm} & \textbf{LLM-based Alignment (Generative Profile)} & \textbf{KGBridge (Disentangled Prompt)} \\ \midrule
			\textbf{Representation} & \textbf{Holistic Summarization:} & \textbf{Disentangled Soft Prompts:} \\
			\textbf{Mechanism} & Compresses diverse history into a single text profile: \newline \textit{``You seem to enjoy action-packed and entertaining content...''} & Continuous relation prompts modeling both domain-shared (e.g., genre patterns) and domain-specific relational semantics. \\ \midrule
			\textbf{Top-3 Recs} & Semantically concentrated recommendations: \newline 1. \textit{Die Hard} \textcolor{red}{(Action)} \newline 2. \textit{Lethal Weapon} \textcolor{red}{(Action)} \newline 3. \textit{The Matrix} \textcolor{red}{(Action/Sci-Fi)} & Semantically diversified recommendations: \newline 1. \textit{Reservoir Dogs} \textcolor{blue}{(Crime/Thriller)} \newline 2. \textit{Spirited Away} \textcolor{blue}{(Animation/Fantasy)} \newline 3. \textit{Walk the Line} \textcolor{blue}{(Drama/Biography)} \\ \bottomrule
		\end{tabular}
	}
\end{table}

	\paragraph{Analysis of Semantic Generalization in LLM4CDSR}	
	User~61353 has interacted with items from multiple genres in the movie domain, including the neo-noir film \textit{Pulp Fiction}, the biographical drama \textit{Ray}, and the fantasy animation \textit{Howl's Moving Castle}, as listed in Table~\ref{tab:case_user}. Under non-overlapping settings, LLM4CDSR constructs a unified textual profile to summarize these heterogeneous behaviors, generating the description: ``You seem to enjoy action-packed and entertaining content, often with a mix of humor and drama...''.
	
	While this representation is semantically coherent, it tends to compress fine-grained preferences into coarse-grained textual themes, a phenomenon we term as \textit{semantic generalization}. As shown in prior studies on transferable recommendation~\cite{hou2022learning}, such holistic textual summarization is sensitive to domain-specific vocabulary shifts, which may weaken precise alignment in downstream recommendation. As a result, although high-level themes are preserved, fine-grained relational cues and preference structures may be diluted, causing the generated recommendations to overemphasize dominant semantic themes while underrepresenting fine-grained or diverse user interests.
	
	To further examine the prevalence of such semantic heterogeneity, we analyze user interaction diversity on the FB~Movie-Book dataset using broad KG-derived genres (e.g., Action, Drama, Sci-Fi, Animation). 
	As illustrated in Fig.~\ref{fig:case_study}, each user interacts with items spanning 11.2 major genres on average, while 98.85\% of users interact with items covering at least four distinct genres. These findings suggest that semantically diverse interaction histories are common in real-world cross-domain recommendation, highlighting the importance of preserving fine-grained semantics under non-overlapping settings.
	
	\paragraph{Relational Soft Prompts as Semantic Controllers}
	In contrast, KGBridge encodes relation-level semantics into soft prompts derived from structured KG relations (e.g., \texttt{subgenre}, \texttt{is\_adaptation\_of}). To capture both transferable and domain-specific knowledge, we construct dual prompt banks consisting of domain-shared and domain-specific prompts, which jointly model cross-domain invariance and domain uniqueness. 	These prompts preserve stable relational patterns and provide structured semantic priors for sequential preference modeling.	By maintaining structural consistency at the relation level while explicitly disentangling shared and domain-specific semantics, KGBridge enables more precise alignment of fine-grained preferences across domains and avoids over-compression of user interests.
	
	Beyond the empirical comparison, these observations also reveal broader implications for future cross-domain recommendation research: (1) \textit{Structural Regularization for LLMs}: Future work could explore hybrid architectures where knowledge graphs serve as structural priors to regularize the generative space of LLMs, thereby mitigating information loss during persona summarization. (2) \textit{Disentangled Knowledge Prompting}: Rather than employing a monolithic profile, adopting disentangled, knowledge-guided prompts can better capture the complex, multi-faceted nature of user preferences across diverse domain boundaries.
	
	\section{Related Work}
	\subsection{KG-enhanced Recommendation}
	Knowledge graphs (KGs) have been widely incorporated into recommender systems to enhance semantic representation, relational reasoning, and interpretability \cite{GuoZQZXXH22}. Existing KG-enhanced recommendation methods generally exploit entity and relation information through embedding learning, graph propagation, attention-based aggregation, or memory-enhanced semantic modeling, enabling richer semantic context for user--item interaction modeling \cite{ZhangYLXM16, WangZXG18, HuangZDWC18, WangZWZLXG18, Wang00LC19, XianFMMZ19, YuRSSKGNH13}. These methods enrich user and item representations with structured semantic dependencies beyond direct interaction signals, thereby improving semantic understanding and recommendation interpretability.
	
	Despite their effectiveness, most existing KG-enhanced recommenders remain confined to single-domain settings and predominantly rely on entity-centric semantic modeling. Since real-world KGs often exhibit severe long-tail structural imbalance, representations derived from sparse or low-frequency entities are often biased toward high-frequency structures and may exhibit limited robustness under sparse semantic propagation. Such limitations become more pronounced in cross-domain scenarios, where semantic transfer must remain robust across heterogeneous domains with varying structural characteristics.
	
	\subsection{KG-enhanced Cross-domain Recommendation}
	To alleviate sparsity and improve transferability across heterogeneous domains, cross-domain recommendation (CDR) has emerged as an important recommendation paradigm. Early CDR methods \cite{ManSJC17, KangHLY19, ZhuGZXXZL021, LiuLLP20} typically follow a representation learning and alignment framework, where domain-specific representations are first learned using collaborative filtering models and then aligned through mapping functions based on overlapping users or shared interaction patterns. More recent approaches further extend this paradigm by employing hierarchical gating \cite{WangZWMZZYR25} or mutual knowledge distillation across multi-view graphs \cite{ZangFCZYZ25}. However, these methods mainly rely on behavioral interactions for domain alignment and generally overlook the structured semantic knowledge encoded in KGs.
	
	To enhance semantic transferability, researchers have further explored KGs as external knowledge sources for CDR. Early attempts incorporate KG-derived semantic relatedness as auxiliary regularization in matrix factorization models \cite{Fernandez-Tobias19}, improving generalization but without explicitly modeling structural semantic dependencies. Subsequent works further exploit KG semantics or structure for transfer, such as leveraging attribute-level knowledge to refine cross-domain preference modeling \cite{WangXTBLL25} or aggregating structurally related entities via preference-aware attention over cross-domain KGs \cite{Li0L23}. More recent approaches jointly model semantic correlations and structural dependencies \cite{WangXTLYL24, LiHL24}, thereby enhancing cross-domain semantic alignment and transferability.
	
	Despite these advances, most existing KG-enhanced CDR methods still rely heavily on entity-centric semantic representations, which remain sensitive to KG sparsity and long-tail structural imbalance. Moreover, many approaches depend on overlapping users or behavioral correspondence for cross-domain alignment, limiting their applicability under privacy-preserving non-overlapping settings.
	
	\subsection{LLM-based Cross-domain Recommendation}
	Beyond KG-enhanced paradigms, recent studies have explored Large Language Models (LLMs) to bridge item semantics across domains by leveraging broad world knowledge~\cite{00020W0Z0LH025, HadadRESR25, WangZL25, GuoLYNY24}. For instance, Liu et al.~\cite{00020W0Z0LH025} leverage LLMs as semantic bridging anchors, while Guo et al.~\cite{GuoLYNY24} propose prompt-enhanced federated content learning. These approaches demonstrate the potential of LLMs in capturing transferable semantic priors across heterogeneous domains.
	
	However, applying generative semantic modeling to recommendation scenarios also introduces practical challenges, including substantial computational overhead and the risk of generating semantically plausible yet weakly grounded associations~\cite{LinDXLCZLWLZGYTZ25}. Such issues become more pronounced in non-overlapping settings, where the absence of shared users weakens the collaborative constraints available for semantic calibration across domains.
	
	Motivated by these limitations, our work revisits the role of KGs in non-overlapping cross-domain recommendation from a relation-centric perspective. Unlike existing methods that primarily depend on entity-level representations or behavioral alignment, KGBridge encodes relation-level semantics as adaptive soft prompts to establish structured semantic anchors across domains. This design enables robust and computationally efficient knowledge transfer while reducing sensitivity to KG sparsity, popularity bias, and the absence of overlapping users.
	
	\section{Conclusion and Future Work}
	
	This paper presented KGBridge, a knowledge-guided prompt learning framework for non-overlapping cross-domain sequential recommendation. Unlike conventional KG-enhanced recommenders that primarily rely on entity-centric representations and overlapping user alignment, KGBridge revisits the role of knowledge graphs (KGs) from a relation-level perspective, treating KG relations as transferable semantic priors for cross-domain recommendation. By encoding relation semantics into soft prompts, the proposed framework dynamically guides sequential representation learning while reducing sensitivity to KG sparsity and popularity bias.
	
	To achieve robust knowledge transfer under non-overlapping settings, we further design a KG-enhanced prompt encoder with correspondence-driven disentanglement to explicitly separate domain-shared and domain-specific semantics. In addition, a two-stage training paradigm combining cross-domain pretraining and privacy-preserving fine-tuning enables stable adaptation without requiring identity-level alignment across domains. Extensive experiments on multiple benchmark scenarios demonstrate that KGBridge consistently outperforms state-of-the-art methods under multiple cross-domain settings and maintains strong robustness under varying levels of KG sparsity. Additional analyses further verify the effectiveness of relation-guided prompting, disentangled semantic modeling, and staged adaptation mechanisms.
	
	Beyond empirical performance improvements, this work highlights the potential of relation-level semantic prompting as a lightweight yet effective alternative to static entity-centric knowledge transfer. Our findings suggest that preserving structural semantic regularities is particularly important for non-overlapping recommendation scenarios, where direct collaborative correspondence is unavailable and semantic alignment becomes increasingly dependent on transferable relational priors. Nevertheless, because the current evaluation remains based on
	DBpedia-aligned KGs, generalization across heterogeneous KG schemas or
	substantially weaker relation overlap requires further investigation.
	
	Future work will further investigate adaptive and controllable knowledge transfer mechanisms for non-overlapping cross-domain recommendation. As demonstrated in our prompt aggregation analysis, the optimal aggregation strategy may vary across domain pairs due to differences in semantic gap and KG structural characteristics. Therefore, we plan to explore adaptive prompt routing and dynamic aggregation mechanisms to better capture heterogeneous cross-domain semantics. In addition, we aim to investigate tighter integration between large
	language models and structured KGs, where relation-guided structural
	priors may improve semantic grounding and enable more controllable
	cross-domain knowledge transfer.
	
	\begin{acks}
			The authors Yuhan Wang, Qing Xie, Mengzi Tang, Lin Li, and Yongjian Liu are supported partially by National Natural Science Foundation of China (Grant No. 62572365) and National Key Research and Development Program of China (2024YFF0907002).
	\end{acks}
	
	\bibliographystyle{ACM-Reference-Format}
	\bibliography{KGBridge-tois}
\end{document}